\documentclass{aa}  
\usepackage{graphicx}
\usepackage[varg]{txfonts}
\usepackage{natbib}
\usepackage{hyperref}

\newcommand\un[1]{{\,\rm #1}}
\newcommand\E[1]{\times10^{#1}}
\newcommand\rs[1]{_\mathrm{#1}}

\newcommand\g{$\gamma$}

\usepackage{xspace} 
\newcommand\tcb{T~CrB\xspace}

\usepackage{ulem} 
\usepackage[dvipsnames]{xcolor}

\begin{document}

\title{Radio component of
the multi-messenger emission\\ from the anticipated T~Coronae Borealis outburst}
\titlerunning{Radio emission from T Coronae Borealis}
\author{O. Petruk\inst{1,2}
    \and
    T. Kuzyo\inst{2}
	\and
	S. Orlando\inst{1}
    \and
    L. Chomiuk\inst{3}
	\and
    F. Bocchino\inst{1}
    \and
    M. Miceli\inst{1,4}
    \and
    V. Beshley\inst{2}
}

\institute{INAF - Osservatorio Astronomico di Palermo, Piazza del Parlamento 1, 90134 Palermo, Italy\\
    \email{oleh.petruk@inaf.it}	
	\and
	Institute for Applied Problems in Mechanics and Mathematics, National Academy of Sciences of Ukraine, Naukova St. 3-b, 79060 Lviv, Ukraine
    \and
    Center for Data Intensive and Time Domain Astronomy, Department of Physics and Astronomy, Michigan State University, East Lansing, MI 48824, USA
    \and
    Dipartimento di Fisica e Chimica E. Segrè, Università degli Studi di Palermo, Piazza del Parlamento 1, 90134, Palermo, Italy
}

\date{Received ...; accepted ...}

\abstract{T~Coronae Borealis is the nearest symbiotic recurrent nova and is expected to undergo a new eruption soon, offering a rare opportunity to study shock evolution and particle acceleration in a dense binary environment.
We predict the radio emission from the anticipated outburst and assess how radio observations can constrain the circumbinary medium (CBM) and cosmic-ray acceleration.
We post-process three-dimensional (3D) hydrodynamic nova simulations with a multi-zone model for particle acceleration and radio emission, computing thermal free-free and non-thermal synchrotron radiation, including the Razin effect, free-free and synchrotron self-absorption for different CBM configurations. 
We also synthesize radio images for comparison with the observed morphology once the source becomes spatially resolvable. Radio morphology during the first weeks should not be interpreted directly as tracing the intrinsic shock geometry, because free-free absorption selectively obscures different regions of the 3D emitting volume.
The intrinsic radio emission is dominated by synchrotron radiation from electrons accelerated at the forward shock, with unabsorbed fluxes reaching $10^2$-$10^3\un{Jy}$ at GHz frequencies during the first day. 
Dense ejecta and shocked CBM material, however, strongly absorb the emission, reducing the observable flux by several orders of magnitude. 
The source becomes largely transparent at GHz frequencies within about one month. 
Radio emission is predicted to remain predominantly non-thermal during the first year, while the thermal component becomes important only for models with particularly dense equatorial enhancements. The radio spectral index evolves from positive values caused by absorption to the optically thin synchrotron value of $-0.5$ after several weeks. A short-lived polarized signal may arise during the first few days if the red giant’s magnetic field significantly contributes to the cosmic-ray-generated field. Early multi-frequency radio monitoring and polarimetry of the forthcoming \tcb eruption will probe the density structure and magnetic field, providing a powerful test of particle acceleration in symbiotic recurrent novae.}


\maketitle

\section{Introduction}
\label{tcb-radio:sec-intro}

Symbiotic recurrent novae are binary systems composed of a white dwarf (WD) accreting from the wind of a giant companion, wherein multiple nova eruptions have been observed. To date, four are known in our Milky Way Galaxy \citep{2010ApJS..187..275S}: RS Oph, V745 Sco, V3890 Sgr, and the subject of this paper: \tcb. 
There are other novae that resemble the symbiotic recurrents (V407 Cyg, V1534 Sco, etc), though with an outburst interval exceeding a century. They have only been observed to erupt once but they have evolved companions and exhibit signatures of shock propagation through a dense stellar wind, typical of the recurrent nova class \citep[e.g.][]{2011A&A...527A..98S}. 

These systems produce energetic shocks when the nova ejecta crash into circumbinary material (CBM), which originates from the red giant (RG) wind and is shaped by the joint interaction of the RG and WD \citep{Sokoloski+06,Delgado&Hernanz19, Page+20}. 
The shocked ambient medium and ejecta produce X-rays, which were first detected in 1985 from RS Ophiuchi \citep[e.g.][]{2006ApJ...652..629B}. The same outburst is known for the first detection of radio emission from a recurrent nova \citep{1985Natur.315..306P}.

In recent years, novae in these systems have also become well-established as accelerators of cosmic rays (CRs), with detections of $\gamma$-rays (from GeV to TeV energies; \citealt{2010Sci...329..817A,2018A&A...609A.120F,2022ApJ...935...44C,2022NatAs...6..689A,2022Sci...376...77H,Molina+26}), and radio synchrotron emission \citep{Hjellming+86, Taylor+89, Nyamai+23, 2024MNRAS.534.1227M, Molina+26}. The next eruption of the nearest symbiotic recurrent nova, \tcb, is expected imminently, and so our collaboration has embarked on a campaign to simulate what the eruption might look like \citep[][hereafter Paper I]{2025A&A...704A.144O}, and to predict signatures in the thermal X-rays (Paper I), $\gamma$-rays and neutrinos \citep[][hereafter Paper II]{2026A&A...711A..96P}, and in this paper, at radio wavelengths. Some of our models predict an eventual neutrino signal; therefore, the \tcb outburst has the potential to be a multi-messenger event.

The highly anticipated outburst of the closest and brightest nova \tcb is a unique once-a-century opportunity to test models of recurrent novae, the structure of the CBM, and the theory of CR acceleration.
In particular, radio observations can provide a complementary probe of the forward shock because the synchrotron emissivity depends on both the population of accelerated electrons and the post-shock magnetic field, while the observed emission is strongly modulated by free-free absorption expected to be present in the dense CBM. Multi-frequency spectra can therefore help disentangle particle acceleration, magnetic-field amplification, and the three-dimensional density structure.

\section{Methodology}
\label{tcb-radio:sect-approach}

\subsection{Hydrodynamic model and CR acceleration}

The present paper is based on the data of 3D numerical HD simulations reported in Paper I. 
After the thermonuclear runaway, the shock expands into the complex circumbinary medium (CBM) consisting of the spherical stellar wind and the equatorial disk enhancement (EDE, which is a result of the orbital motion of the binary companions), both centered on RG. The spatial density distribution in the first component is proportional to $n\rs{w}\cdot(r/\mathrm{pc})^{-2}$, in the second component to $n\rs{ede}\cdot\exp(-r^2/h^2)$ with scales $h$ different along different Cartesian axes. The third CBM component is the accretion disk, which we model as a disk with a concave parabolic surface having the maximum radius $R\rs{disk}=0.5\un{a.u.}$ and the maximum vertical extent $h\rs{max}=0.3\un{a.u.}$. The density of the accretion disk is modeled as $ n\rs{disk}(\mathbf{r})=10^3\, n\rs{rg}(\mathbf{r})$ where $n\rs{rg}$ is the density provided by the first two components. Paper I provides further details on the models, numerical setup, and simulations and includes an illustration of the initial conditions. 

In the present paper, we consider the same four models of \tcb as in Paper II. Sect.~2.1 and Table~1 in Paper II summarize the parameters of these models. Besides some differences in the EDE length scales $h$, the models differ mainly in the explosion energy $E\rs{bw}$ and the maximum EDE density $n\rs{ede}$. Numerically, $(E\rs{bw},n\rs{ede})$ are $(3.0,0.1)$ in RUN04, $(3.0,10)$ in RUN03, $(10.,1.0)$ in RUN10, $(3.0,0.1)$ in RUN14 with the energy $E\rs{bw}$ measured in $10^{43}\un{erg}$ and the number density $n\rs{ede}$ in $10^{7}\un{cm^{-3}}$. Unlike the other three models, RUN14 does not include an accretion disk. In all models, $n\rs{w}=10^{-3}\un{cm^{-3}}$. RUN04 is our reference model. 

\begin{figure}
  \centering 
  \includegraphics[width=\columnwidth]{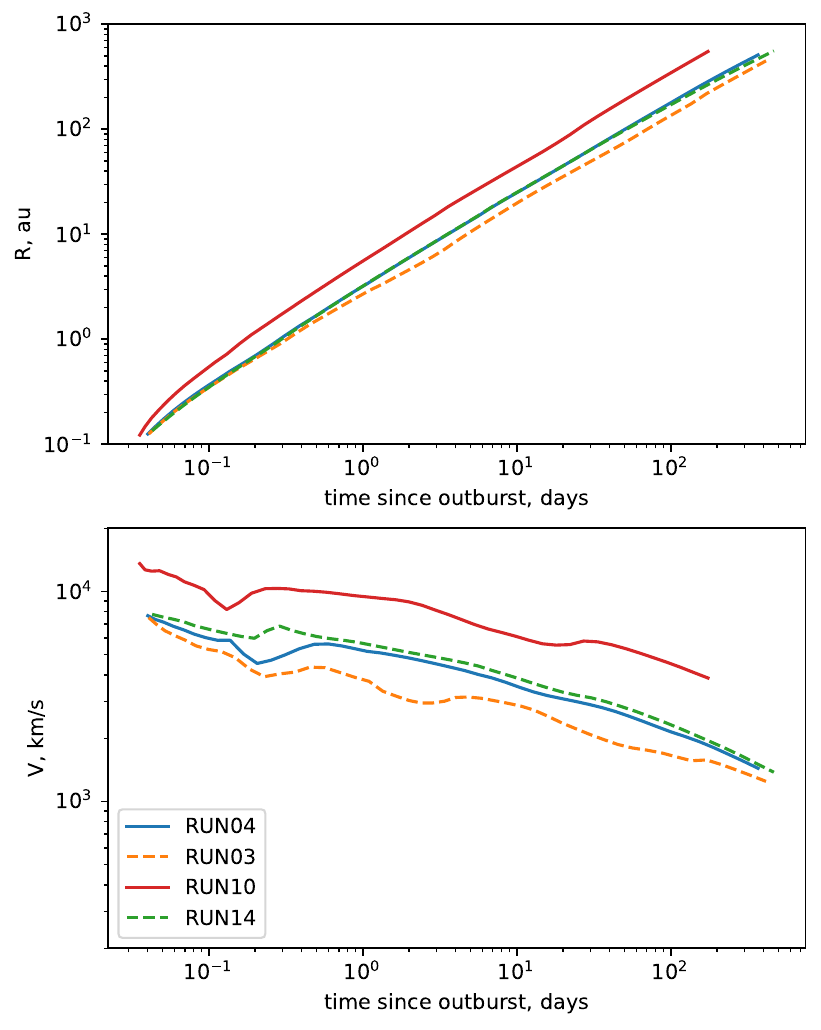} 
  \caption{%
       The evolution of the average shock radius $R$ (top) and the average shock velocity $V$ (bottom) for our \tcb models. The plot shows the average values; $1\sigma$ errors in the direction-averaged values are approximately $0.1R$ and $0.2V$.
  }
  \label{tcb:fig-shock-front-params}
\end{figure}

Fig.~\ref{tcb:fig-shock-front-params} compares the direction-averaged shock radius $R$ and velocity $V$ over the simulation period, highlighting similarities and differences between the models. Notably, the average radius is similar across the models, and the velocity variations are moderate. The evolution of the density cross-sections in the reference model is shown with Animation~D1 in Appendix~\ref{tcb-radio-app-movies}.

The numerical models are purely hydrodynamical. We superimpose the magnetic field (MF) on the 3D HD data by incorporating its spatial distribution as described in Sects.~\ref{tcb-radio:sect-disorderedMFmodel} and \ref{tcb-radio:sect-orderedMFmodel}. 

Relativistic electrons that produce synchrotron emission are accelerated at the forward shock. We compute the energy spectrum of cosmic rays in each cell over the shock surface, thereby employing a multi-zone emission model (see details in Sect.~2 of Paper~II and Appendix~\ref{tcb-radio-app-e-spectrum}).
For radio emission, we consider the power-law part of the electron energy distribution $N(E)dE=KE^{-s}dE$. 
The values we adopted for plots in the present paper are the same as in Paper II, namely, $s=2$, the fraction of the kinetic energy of the shock transferred to CRs $\xi\rs{cr}=0.1$ and the relative normalization of the electron to proton momentum distributions $K\rs{ep}=0.01$ (Appendix~\ref{tcb-radio-app-e-spectrum}).

\subsection{Synthesis of the radio emission}

Two distinct physical processes drive the radio emission in this nova system: thermal free-free continuum and synchrotron radiation from relativistic electrons. Together, these mechanisms produce three distinct observational components in the radio flux: thermal free-free emission from the RG wind (which dominates the radio flux during quiescence), thermal free-free emission from the expanding, warm, photoionized ejecta, and nonthermal synchrotron emission from shock-accelerated electrons. 

Radio synchrotron emission is produced by relativistic electrons in the disordered (Sect.~\ref{tcb-radio:sect-disorderedMFmodel}) and ordered (Sect.~\ref{tcb-radio:sect-orderedMFmodel}) components of MF.
Since no significant radio polarization has been reported in recurrent novae to date  \citep[even when observations were performed in a full-Stokes mode; e.g.][]{2008ApJ...688..559R,2024MNRAS.534.1227M}, we do not track the orientation of magnetic vectors. The thermal radio continuum is calculated as described in Sect.~\ref{tcb-radio:sect-thermal}. 

We stress that the radio spectral index in our models is intrinsically multi-zone: densities, shock velocities, electron acceleration, emission, and absorption vary across different locations. 
Starting from the 3D HD simulations, with the magnetic field prescribed as described in Sect.~\ref{tcb-radio:sect-disorderedMFmodel} below, we calculate emissivity in each numerical cell and then synthesize observables, accounting for absorption along the line of sight (LoS) and orientation of the binary system relative to the observer. 
When generating synthetic observables, we adopt the distance to \tcb $d=890\un{pc}$ \citep{2021AJ....161..147B}. Unless specified otherwise, we fix an orbital inclination of $55\degr$ \citep{2025ApJ...983...76H} and a radio frequency of $1\un{GHz}$. 
As in Papers I and II, we adopt the same arbitrary choice for the orbital phase of the binary system by applying an initial $20\degr$ rotation about the $z$-axis, such that RG lies on the far side of the binary relative to the observer. 
This enables a direct comparison with results from our previous papers.

\begin{figure}
  \centering 
  \includegraphics[width=\columnwidth]{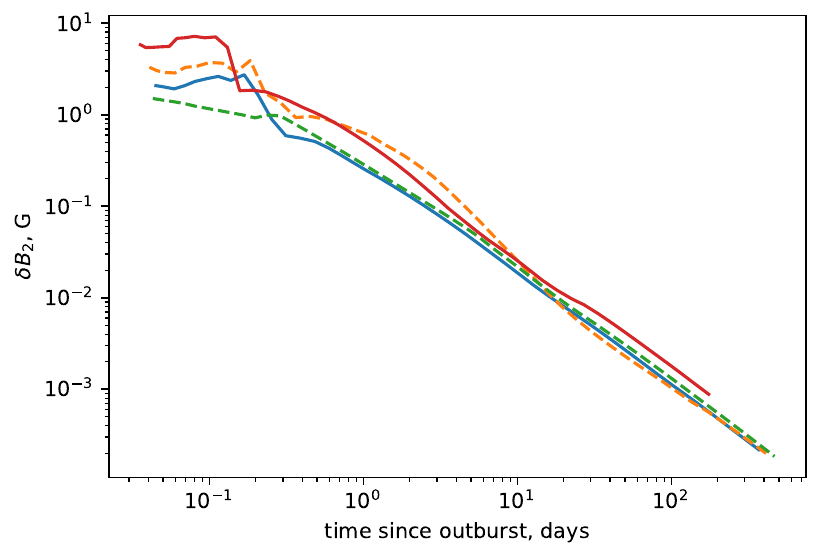} 
  \caption{%
       The evolution of the immediately post-shock MF strength averaged over the shock surface. The plot shows the average values; the $1\sigma$ errors of the shock-averaged $\delta B$ are approximately $(0.3-0.5)\delta B$ except for the period of the shock propagation across the accretion disk ($t\lesssim0.3\un{day}$) when the spread of the values is larger than the average, as a consequence of huge differences in the density of the disk and other regions.}
  \label{tcb:fig-shock-front-deltaB}
\end{figure}

\subsection{Disordered magnetic field}
\label{tcb-radio:sect-disorderedMFmodel}

In the immediately pre-shock region, the strength of the disordered MF is calculated as the saturated level of the \citet{2004MNRAS.353..550B} non-resonant instability with the magnetic energy density 
\begin{equation}
 \frac{\delta B_1^2}{4\pi}\simeq\xi\rs{cr}\rho\rs{1}V^3/c
 \label{tcorbor:eq-dB}
\end{equation}
where the subscript `1' marks the pre-shock values.
Fig.~\ref{tcb:fig-shock-front-deltaB} shows the temporal variation of the shock-averaged strength of this MF derived from our models. The initial plateau results from the shock propagation in the dense accretion disk. 

The orientation of $\delta B$ is assumed to be random and isotropic, and, therefore, 
the average of each of the three components is $\langle\delta B_{r,\theta,\phi}^2\rangle=\delta B^2/3$. The evolution of the normal and tangential components of a magnetic vector $\mathbf{B}$ may be described in the Lagrangian approach by preserving the magnetic flux in a given sector \citep{1974ApJ...188..501C,1998ApJ...493..375R,2023MNRAS.518.6377P}: 
\begin{equation}
 B\rs{\|}(a,t)=B\rs{1\|}(a)\left(\frac{a}{r}\right)^2,\quad 
 B\rs{\perp}(a,t)=B\rs{1\perp}(a)\frac{\rho(a,t)}{\rho\rs{1}(a)}\frac{r}{a}
 \label{tcorbor:eq-Bdownstream}
\end{equation}
where $a$ is the Lagrangian coordinate and the index `1' marks the pre-shock values. 
Therefore, the disordered MF 
\begin{equation}
 \delta B\equiv \sqrt{\left\langle\delta B\rs{r}^2\right\rangle+\left\langle\delta B\rs{\theta}^2\right\rangle+\left\langle\delta B\rs{\phi}^2\right\rangle}
\end{equation}
evolves in a given fluid element as 
\begin{equation}
 \delta B_2(a,t) = 
 \delta B_1(a,t\rs{i})\left(
 \frac{1}{3}\left[\frac{a^2}{r^2}\right]^2+
 \frac{2}{3}\left[\frac{\rho(a,t)}{\rho\rs{o}(a)}\frac{r}{a}\right]^2
 \right)^{1/2}
 \label{tcbsynch:dBevol}
\end{equation}
where $t\rs{i}$ is the time when this fluid element was shocked, subscript `2' marks the downstream values, $\delta B_1(a,t\rs{i})$ is calculated with equation~(\ref{tcorbor:eq-dB}) with the Lagrangian shock velocity $V(a,t\rs{i})$. 
This expression results, in particular, in $\delta B_2/\delta B_1=\sqrt{11}$ at the shock with the compression factor $\rho_2/\rho_1=4$.  

The length scale for the wave damping is larger than the nova’s radius (Appendix~\ref{tcb-radio:sect-app-dBevol}). Therefore, we consider that the evolution of $\delta B$ is due only to changes in the HD structure of the flow. 
Since this component of MF is generated by the forward shock, 
the formula (\ref{tcbsynch:dBevol}) may be applied only to fluid elements corresponding to the shocked ambient medium.
We exclude the WD and RG interior from the calculations of $\delta B$ because the shock does not penetrate into the dense star interior.

\begin{figure}
  \centering 
  \includegraphics[width=\columnwidth]{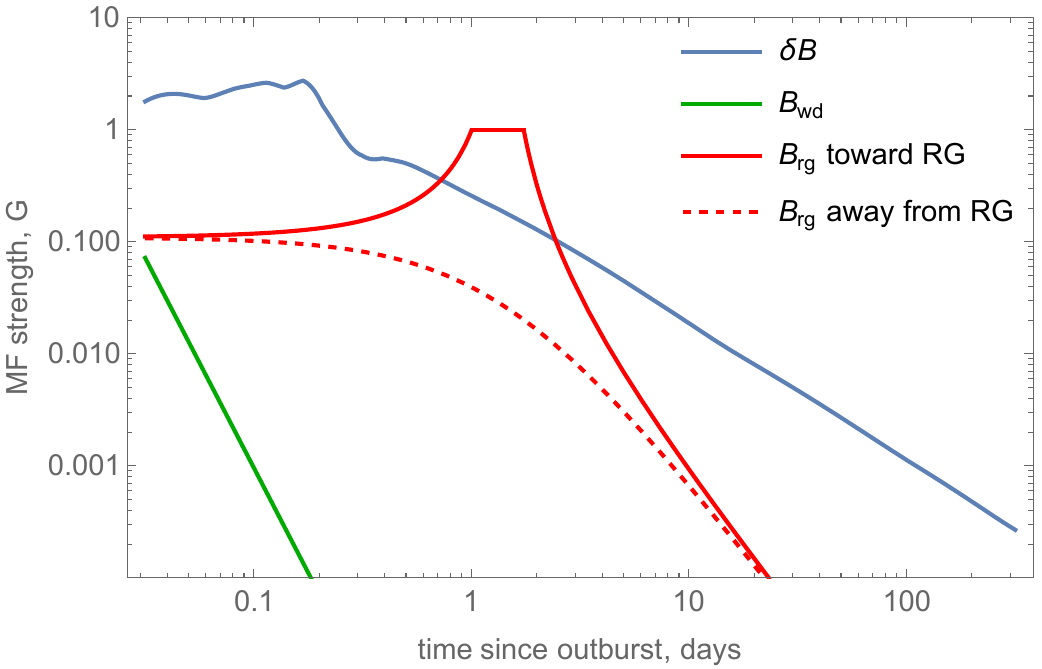} 
  \caption{%
       Comparison of the MF components for the model RUN04. The        
       disordered MF averaged over the shock surface is shown with the blue line. The ordered components are given by equation~(\ref{tcb:MFordered-estimates}): green line for WD and the red lines for RG. 
       In order to convert distance to time, we used the evolution of the average shock radius $R(t)$ from Fig.~\ref{tcb:fig-shock-front-params}. The solid red line corresponds to the portion of the shock moving toward the RG, and the dashed one to the opposite direction. The strength is for the immediately post-shock MF, i.e., after the shock compression, which is unity for the red lines (the shock is parallel) and four for the green line (the shock is perpendicular almost everywhere). 
       The plateau in the solid red line on the day 2 corresponds to the RG diameter.}
  \label{tcb:fig-synchrotron-MFordered}
\end{figure}

\subsection{Circumbinary magnetic field}
\label{tcb-radio:sect-orderedMFmodel}

In our models of \tcb, the disordered MF generated at the forward shock produces high synchrotron fluxes. Both stars in the binary system have their own MFs. 
In order to estimate the effect of these ordered MF components, we consider the dipole MF for WD and the \citet{1958ApJ...128..664P} model with a large rotation period for RG: 
\begin{equation}
 B\rs{wd}\simeq \frac{B\rs{wd}'R\rs{wd}^3}{r^3},\qquad 
 B\rs{rg}\simeq \frac{B\rs{rg}'R\rs{rg}^2}{r^2}
 \label{tcb:MFordered-estimates}
\end{equation}
where $r$ is the distance from the star. 
On the surface of RG, the MF strength could be of the order $1-10\un{G}$ \citep{2015A&A...574A..90A}.  
Some symbiotic stars exhibit magnetic properties similar to intermediate polars, with the MF strengths of 0.1-10 MG \citep{2008MNRAS.387.1157R}. However, the absence of such variability in \tcb points to a weaker magnetic field in the WD. Furthermore, because a highly magnetized WD would suppress the formation of an accretion disk, confirming the disk's existence during the upcoming nova event would definitively signal a low magnetic field in the WD in the \tcb system. 

Fig.~\ref{tcb:fig-synchrotron-MFordered} compares the disordered MF to these ordered components at the forward shock in the RUN04 model. 
To produce the figure, we take $B\rs{wd}'=1\un{MG}$ at $R\rs{wd}=0.0045R_\odot$ and $B\rs{rg}'=1\un{G}$ with radius $R\rs{rg}=64R_\odot$; the orbital separation is $b=0.9\un{AU}$. 
We see from the figure that the MF of the WD is inefficient in modifying the synchrotron flux from \tcb nova. Instead, the RG's magnetic field could become important during the second day, when the ejected material flows around the RG star. During this period, a polarized radio signal could potentially emerge, provided that thermal emission is not dominant and absorption does not suppress the polarized signal. If $B\rs{rg}$ is higher, the red lines shift upward in the plot. Consequently, the period during which a polarized signal may be expected could extend into the first few days after the eruption, when the interaction region still occupies a large fraction of the shock surface. This makes polarimetric observations during the first days particularly desirable, as they could provide important information about the magnetic field in the \tcb system. At later times, the fraction of polarized radio emission should be negligible, consistent with what is generally observed in novae. 

If a polarized signal emerges in observations of the \tcb nova, then detailed modeling of the MF in the binary system should be performed.
In the present paper, we instead follow a simplified approach. We consider the RG field $B\rs{rg}$ as the dominant in the CBM.\footnote{In the approximation (\ref{tcb:MFordered-estimates}) which we use, the Parker spiral geometry is not followed; we capture only the radial decline of the field strength.} After the shock passage, it follows the first relation from (\ref{tcorbor:eq-Bdownstream}): 
\begin{equation}
 B_2(\vec a,t)\simeq\frac{B\rs{rg}'R\rs{rg}^2}{|\vec a-\vec b|^2}\ \frac{a^2}{r^2}
 =\frac{B\rs{rg}'R\rs{rg}^2}{a^2+b^2-2a_xb}\ \frac{a^2}{r^2}
\end{equation}
where $a$ is the absolute value of $\vec a=(a_x,a_y,a_z)$, $\vec b=(b,0,0)$, $a$, $b$, $r$ are measured from WD (the rightmost relation holds for the system configuration when RG is located on the $x$-axis).
In this approach, we do not trace the orientation of the magnetic vectors and take $\sqrt{2/3}\,B_2$ for the component perpendicular to the line of sight. At each point, the total MF is the sum ${\cal B}_2=\left(\delta B_2^2+B_2^2\right)^{1/2}$.

\subsection{Unpolarized radio synchrotron emission}
\label{tcb-radio:sect-disorderedMF}

We calculate the radio emission using a simplified approach by considering only the disordered MF. Since radio polarized emission is not detected in recurrent novae, we limit the analysis by focusing only on the Stokes parameter $I$. 

The local Stokes ${\cal I}$ (emissivity) is given by the analytical formula (28) in \citet{2016MNRAS.459..178B}, which is derived for a power-law electron energy distribution $N(E)dE=KE^{-s}dE$ and a completely random MF: 
\begin{equation}
  {\cal I} = c_1 {\cal A}(s) \left(\frac{\nu}{c_2}\right)^{-(s-1)/2}  
  K\,{\cal B}_\perp^{(s+1)/2}\, 
  \Gamma\left(\frac{s+5}{4}\right)
  \label{tcb:Isynchr}
\end{equation} 
where ${\cal I}$ is in units $\mathrm{{erg}\,{cm^{-3}\,s^{-1}\,Hz^{-1}}}$, ${\cal B}_\perp\equiv\sqrt{2/3}\,{\cal B}_2$ is an effective MF perpendicular to the LoS, $\nu$ the radio frequency, $c_1=\sqrt{3}e^3/(m\rs{e}c^2)=2.35\E{-22}\un{cgs}$, $c_2=3e/(2\pi m\rs{e}^3c^5)=1.25\E{19}\un{cgs}$, 
\begin{equation}
 {\cal A}(s) = \frac{1}{4}\frac{s+7/3}{s+1}
    \Gamma\left(\frac{s}{4}+\frac{7}{12}\right)
    \Gamma\left(\frac{s}{4}-\frac{1}{12}\right)
\end{equation}
that is ${\cal A}(2)=0.736$ for $s=2$. 

For the adopted $s=2$, equation~(\ref{tcb:Isynchr}) gives 
\begin{equation}
  {\cal I} = 4.0\E{-13}\,
  \nu^{-1/2}\,K\,{\cal B}^{3/2}
  \un{\frac{erg}{cm^{3}\,s\,Hz}}.
\end{equation} 
Thus, the radio synchrotron intensity depends both the normalization of the accelerated-electron population and the magnetic-field strength. In the present model these quantities are coupled through the adopted prescriptions for particle acceleration and magnetic-field amplification.

\subsection{Thermal free-free radio emission}
\label{tcb-radio:sect-thermal}

The thermal free-free emissivity of cosmic plasma with the total number density $n$ and temperature $T$ at a frequency $\nu$ is \citep[][eq.3.54]{Spitzer1998}:
\begin{equation}
j_{\nu} = 6.84 \times 10^{-38}\ \frac{g\rs{ff} Z\rs{i}^2 n\rs{e} n\rs{i}}{T^{1/2}}\ e^{-h \nu / kT} \ 
\un{{erg}\,{cm^{-3}\,s^{-1}\,Hz^{-1}}}
\label{tcb:thermal-emmisivity}
\end{equation}
with $Z\rs{i}\approx1.2$, $n\rs{e} n\rs{i}\approx 0.25 n^2$ and the Gaunt factor at radio frequencies \citep[][eq.3.55]{Spitzer1998}:
\begin{equation}
g\rs{ff} \approx 9.77 \left[1 + 0.130\ \ln \left(\frac{T^{3/2}}{Z\rs{i}\nu}+e^{-6.9}\right)\right]
\end{equation}
where we added a term $e^{-6.9}$ which limits the minimum value of the Gaunt factor to  $g\rs{ff}\approx 1$ as it occurs either at high frequencies (e.g., in the optical range) or at low temperatures \citep[e.g.][]{2014MNRAS.444..420V} when the first term inside the logarithm vanishes. Below $T\approx 10^4\un{K}$, rapid recombination causes the plasma ionization fraction to drop sharply, suppressing thermal bremsstrahlung. To account for this effect, we set $j_{\nu}=0$ for cells with $T<10^{3.9}\un{K}$. 
The post-shock temperature is calculated self-consistently from pressure and density in our numerical data. 

\begin{table*}
\caption{Radio fluxes at frequency $\nu$ from our models of \tcb in a quiescent state.}
\begin{tabular}{cc|c|cc|cc|c}
\hline
Model & accretion & $\nu$ & unabsorbed & fractional contributions & absorbed & fractional contributions & radio index\\
      & disk & GHz & flux, Jy & from wind / EDE / disk & flux, mJy & from wind / EDE / disk & 5-45 GHz\\
\hline
RUN04 & Y & 45  & 0.738  & 0.0001 / 0.0000 / 0.9999  & 0.0318 & 0.9975 / 0.0025 / 0.0000 & 0.99 \\
RUN04 & Y & 9.0 & 0.802  & 0.0001 / 0.0000 / 0.9999  & 0.0076 & 0.9915 / 0.0085 / 0.0000 & -- \\
RUN04 & Y & 5.0 & 0.861  & 0.0001 / 0.0000 / 0.9999  & 0.0036 & 0.9862 / 0.0138 / 0.0000 & -- \\
RUN04 & Y & 1.0 & 1.271  & 0.0001 / 0.0000 / 0.9999  & 0.0002 & 0.9665 / 0.0335 / 0.0000 & -- \\
RUN03 & Y & 45  & 3.171  & 0.0000 / 0.0003 / 0.9997  & 0.3530 & 0.0231 / 0.9769 / 0.0000 & 2.0 \\
RUN10 & Y & 45  & 0.874  & 0.0001 / 0.0000 / 0.9999  & 0.0357 & 0.8332 / 0.1668 / 0.0000 & 1.1 \\
RUN14 & N & 45  & $8.43\E{-5}$ & 0.9990 / 0.0010 / 0.0000  & 0.0339 & 0.9976 / 0.0024 / 0.0000 & 1.0 \\
\hline
\end{tabular}
\tablefoot{The fluxes are calculated as thermal free-free emission with $T=10^4\un{K}$ from the pre-outburst density structures contained within a cube of side 2.56 AU after the system has been rotated to match the orbital inclination.} 
\label{tcb:table-quietfluxes-HDmodels}
\end{table*}

\subsection{Absorption and suppression of emission}
\label{tcb-radio:absorp-disorderedMF}

Radio emission is subject to absorption. It can be accounted for as 
$I=I_0\exp(-\tau)$, where $I$ and $I_0$ are the final and the initial intensity, and the optical depth $\tau(\nu)$ at a given frequency $\nu$ is given by an integral of the absorption coefficient $\mu$ along the LoS. At radio wavelengths, the observed emission may be modified by thermal free-free absorption and synchrotron self-absorption: $\mu=\mu\rs{ff}+\mu\rs{ssa}$. 

The free-free absorption coefficient at radio wavelengths is \citep[][eq.~3.57]{Spitzer1998}
\begin{equation}
\mu\rs{ff} = 0.0177 
  \ \frac{g\rs{ff}Z\rs{i}^2 n\rs{e} n\rs{i}}{T^{3/2} \nu^2}\ \un{cm^{-1}}.
 \label{tcorbor:eq-muff}
\end{equation}
It describes the absorption of radio radiation by the thermal ionized plasma and therefore applies to any radio photons traversing the absorbing medium, irrespective of whether they originate from thermal bremsstrahlung or synchrotron emission. 
The value of the coefficient $\mu\rs{ff}\sim 10^{-19}n^2T^{-3/2}\un{cm^{-1}}$ for $\nu=1\un{GHz}$. Then, the depth $\tau\rs{ff}\simeq\mu\rs{ff}R\sim 10^{-12}n^2$ in the warm plasma with $T=10^4\un{K}$ along the distance $R\sim1\un{AU}$. This implies that, in order for the plasma to be opaque, the density should be $n\gtrsim10^{6}\un{cm^{-3}}$. In our models, such a density could be in the shocked CBM $n\rs{ede}\sim 10^7\un{cm^{-3}}$ and in the ejecta $n\rs{ej}\sim 10^{10}\un{cm^{-3}}$. At the early times, the accretion disk adds a factor $10^3$ to the CBM density. Therefore, as is commonly expected, the absorption of radio emission by the thermal plasma in \tcb should be quite strong.

The synchrotron self-absorption occurs within the same population of relativistic electrons responsible for the synchrotron emission. The corresponding absorption coefficient acts on the total specific intensity and therefore affects radio photons of any origin, including thermal free-free emission, when they traverse the synchrotron-self-absorbing region. 
The absorption coefficient is \citep[][eq.~3.52]{1970ranp.book.....P}
\begin{equation}
 \mu\rs{ssa}=c_6(s)K{\cal B}_\perp^{(s+2)/2} (\nu/c_2)^{-(s+4)/2}
 \ \un{cm^{-1}}
\end{equation}
where $c_6=8.61\E{-41}\un{cgs}$ for $s=2$.
If we use the expression for the normalization $K\simeq 7.8\E{-20}n\rs{cr}T^{1/2}$ \citep[equation~11 in][]{2026A&A...711A..96P} where $n\rs{cr}\sim 10^{-4}n$ is the number density of CRs, we obtain $\mu\rs{ssa}\simeq 1.3\E{-33}nT^{1/2}B^2\un{cm^{-1}}$ that converts into $\tau\rs{ssa}\sim 10^{-16}n$ for the shock temperature $T\sim 10^8\un{K}$, ${\cal B}\sim 1\un{G}$ and $R\sim 1\un{AU}$. Thus, even densities $n\sim 10^7$-$10^{10}\un{cm^{-3}}$ do not make synchrotron self-absorption as significant as free-free absorption. 

These estimates are robust under conditions and over the density range considered in this work, although the exact opacity depends on the structure of the absorbing plasma which we treat in 3D.

To account for radiative transfer, the absorption of radio emission is calculated cell-by-cell along each LoS through the data cube. This allows us to fully capture the local conditions that modify absorption. Specifically, for any given cell, we determine the optical depth toward the observer and integrate the remaining, non-absorbed radiation along the line of sight.

There is another process that can lower the synchrotron flux in novae: the Razin-Tsytovich effect. In contrast to free-free and self-synchrotron absorption, where radio photons are emitted normally and then absorbed on the way to the observer, this is an emission-suppression mechanism. 
When the radiation frequency $\nu$ approaches the plasma frequency $\nu\rs{p}\simeq 9000n\rs{e}^{1/2}\un{Hz}$, the refractive index of plasma $n\rs{rf}=\sqrt{1-\nu\rs{p}^2/\nu^2}$ effectively drops. This destroys relativistic beaming by widening the emission cone and reducing the power of the synchrotron radiation. As a result, synchrotron emission is exponentially suppressed 
${\cal I}'\simeq{\cal I}\cdot\exp\left(-\nu\rs{R}/\nu\right)$ at frequencies below the frequency $\nu\rs{R}\simeq 20n\rs{e}/{\cal B}_\perp\un{Hz}$ \citep{1966MNRAS.131..237H}. At the earliest times, when $n\rs{e}\sim 10^7\un{cm^{-3}}$ and ${\cal B}_\perp\sim 1\un{G}$, the Razin frequency could be $\nu\rs{R}\sim 0.1\un{GHz}$. 
The density is higher in the accretion disk. However, ${\cal B}\propto n^{1/2}$ at the shock (equation \ref{tcorbor:eq-dB}) and ${\cal B}\propto n$ downstream (Equation \ref{tcbsynch:dBevol}); therefore, the cutoff frequency $\nu\rs{R}$ is not expected to be much higher in the accretion disk but could affect the GHz frequencies. 
We expect the Razin frequency to decrease with time. Indeed, $n$ decreases, ${\cal B}\propto n^{1/2}V^{3/2}$, shock decelerates slowly (Fig.~\ref{tcb:fig-shock-front-params}), therefore, approximately, $\nu\rs{R}\propto n^{1/2}$. 
We account for the Razin effect in the present paper and demonstrate its effect on the radio emission from our models of \tcb.

\section{Results}

\subsection{Emission and absorption from the outer CBM}

The unshocked CBM is photoionized by high-energy emission from the nova eruption and the nuclear burning white dwarf, and before the eruption, it is probably ionized by accretion onto the WD.  Expected to have temperature $10^4$ K, this gas absorbs and emits at radio wavelengths. To understand the level of its importance for modeling the nova outburst, we synthesized the radio emission from our models in a quiescent state. It is calculated as thermal free-free radiation from the pre-outburst density structures in our models of the binary system, considering a cubic volume with a specified side length.

The emissivity, being proportional to $n^2$, decreases with distance from the RG and the orbital plane. 
We found that the unabsorbed flux at 45 GHz changes by only $1\%$ when the side length of the cubic volume is increased from 1 to 3 AU (because the dominant contribution comes from the accretion disk, which has a radius 0.5 AU and maximum thickness 0.3 AU). Instead, the absorbed flux increases by about a factor of 2 when we increase the length of the cube side from 1 to 2 AU. By increasing it from 2 to 3 AU, the absorbed flux increases by about $10\%$. Therefore, we chose 2.56 AU as a cube side. At larger distances, the contribution from unshocked CBM to both emission and absorption is negligible even during the quiescent state. The shock reaches that distance about 17 hours after the eruption (Fig.~\ref{tcb:fig-shock-front-params}). Therefore, the unshocked CBM might potentially affect emission from the nova during the first day. 

Radio fluxes and spectral indices from our models in the quiescent state are presented in Table~\ref{tcb:table-quietfluxes-HDmodels}, where both the unabsorbed and absorbed flux densities are shown. There are also the spectral radio indices listed in the table; they were calculated from absorbed fluxes between 5 and 45 GHz. 
A few conclusions follow from Table~\ref{tcb:table-quietfluxes-HDmodels}.
\begin{itemize}
\item Our calculations agree with measurements of the radio fluxes from \tcb \citep{2019ApJ...884....8L,2025A&A...702A.276P}. The observed flux density in a quiescent state at $45\un{GHz}$ is of order $0.1\un{mJy}$ \citep{2025A&A...702A.276P}, while our models with various sets of parameters cover the range of absorbed flux $0.03-0.3\un{mJy}$ at 45 GHz and about 2 orders lower at 1 GHz (due to absorption). The spectral radio index between 5 and 45 GHz is $\alpha=0.98$ in the model RUN04, the same as the average index for seven observations by \citep{2019ApJ...884....8L}.
\item The unabsorbed flux in our models is of the order of $1\un{Jy}$, with the largest value for RUN03, where the density of EDE is 2 orders of magnitude higher than in the basic model RUN04.\footnote{The flux in RUN03 is not 100 times higher than in RUN04 because the density in the disk is also affected by the wind structure, which is the same in all our models.} 
\item In the models with the accretion disk (RUN04, RUN03, RUN10), more than 99\% of the unabsorbed emission rises from this disk. In contrast, the flux after absorption is dominated by the outer layers of the RG wind or, in the RUN03 model, of the EDE. Radio waves from other components can barely escape the opaque medium.
\item In the model RUN14 -- which is the same as RUN04 but without the accretion disk -- the unabsorbed flux is dominated by the RG wind. It is $1.6\E{4}$ times lower than in the RUN04 and RUN10 models where the disk is present. Nevertheless, the absorbed flux is comparable to that in the RUN04 and RUN10 models.
\item The absorption in the CBM in models with the accretion disk is very strong. For example, in the RUN04 model, the ratio of unabsorbed to absorbed fluxes is $4.4\E{4}$ at 45 GHz. It is $1.3\E{7}$ at 1 GHz.
\item In the model RUN14 (without the accretion disk), the ratio of unabsorbed to absorbed fluxes is just $2.5$ at 45 GHz. 
\end{itemize}

Therefore: i) most of the radio emission in the models containing a disk originates from the disk and is subsequently absorbed within the disk; ii) the maximum value of the flux reduction factor by CBM beyond the disk is about 2 or 3. 
The absorption by this medium is higher near the equatorial plane (where the density of EDE is highest) and close to RG (where the wind density is highest), i.e., unshocked RG wind/EDE is not optically thin at radio wavelengths only shortly after the nova outburst. Over time, the absorption of radiowaves in the unshocked CBM decreases because the expanding shock leaves less unshocked CBM with effective absorption. In about 20 hours, the forward shock expands to a radius of 3 AU, beyond which the role of the unshocked CBM is negligible, even in the quiescent state. Instead, the emission is absorbed within the shocked material.

Similarly, the contribution of the unshocked CBM to the emission decreases with time as the shock expands. Beyond the accretion disk, it contributes less than $\sim 10^{-4}\un{Jy}$ to the unabsorbed flux and less than $\sim 10^{-5}\un{Jy}$ to the absorbed flux. As shown later (Figs.~\ref{tcb:fig-synchrotron-flux} and \ref{tcb:fig-synchrotron-flux-abs}), these contributions are negligible compared to the fluxes from the nova. 

In this paper, we neglect the emission and absorption from the unshocked CBM (especially from the accretion disk) which could contribute somehow at the time before shock expands to about 2 AU, i.e. within the first 6 hours after the \tcb nova outburst. If radio observations are available within this period, the role of the unshocked CBM should be re-evaluated, in particular, the absorption in the portion of the accretion disk that remains outside the forward shock.

\subsection{Emission from the nova remnant}

Although the term \textit{nova remnant} is conventionally used for the late stages of nova evolution, in this work we use it to refer to the region enclosed by the forward shock at all times, including the earliest phases after the outburst.

The radio emission from the nova remnant consists of thermal free-free and non-thermal synchrotron components, both of which are heavily absorbed during the early stages of the evolution.
We first focus on the unabsorbed emission. 

It is instructive to examine the time evolution of these components through the radio images. The snapshots for different models of \tcb are presented in Appendix~\ref{tcb-radio-app-images} and the animations for the RUN04 model in Appendix~\ref{tcb-radio-app-movies}. As expected, differences between models are caused by variations in ambient density and explosion energy. In particular, the RUN14 model, which does not contain an accretion disk, has an early morphology quite distinct from the other models. The morphological differences between the thermal (Fig.~\ref{tcb:fig-thermal-radio-maps}, Animation~D4a) and non-thermal (Fig.~\ref{tcb:fig-synchrotron-radio-maps}, Animation~D4a) components are also quite evident. 
Namely, the synchrotron emission rises primarily from the outer regions of the remnant, where the electron acceleration occurs. The synchrotron-emitting layer is rather thin. Instead, the thermal emission originates from deeper locations, reflecting denser regions either in the ejecta or in the swept-up ambient medium (accretion disk and EDE). At the same time, CBM or ejecta heated by the shock to temperatures exceeding $\sim 10^7$ K does not contribute significantly to the thermal radio emission, as indicated by the different spatial extents of the regions responsible for the various emission mechanisms (Fig.~\ref{tcb:fig-thermal-radio-maps} versus Fig.~\ref{tcb:fig-synchrotron-radio-maps}).

\begin{figure}
  \centering 
  \includegraphics[width=\columnwidth]{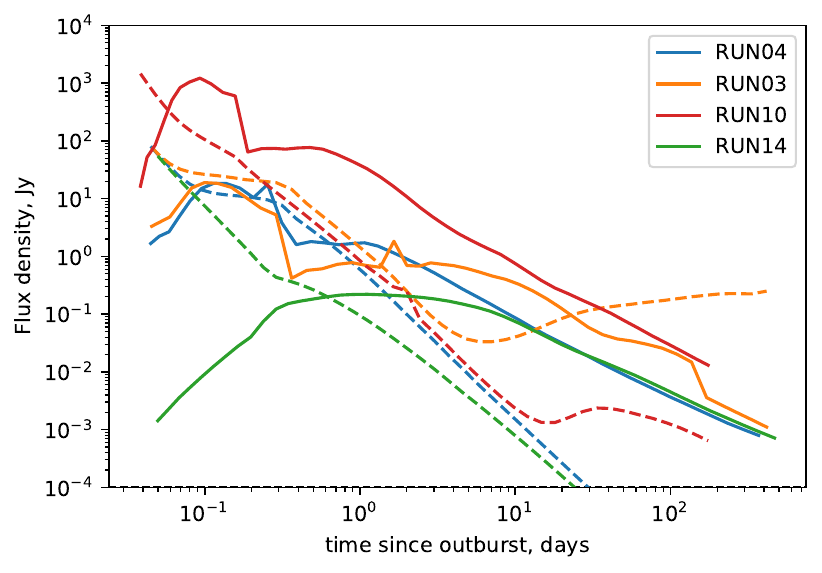} 
  \caption{%
       The evolution of the volume-integrated synchrotron flux with the Razin effect (solid lines) and thermal free-free (dashed lines) flux at $1\un{GHz}$ at the distance of Earth for different \tcb models. Absorption is not included nor is emission from the unshocked ambient medium. 
  }
  \label{tcb:fig-synchrotron-flux}
\end{figure}

Fig.~\ref{tcb:fig-synchrotron-flux} shows the unabsorbed light curves for our models. 
The unabsorbed fluxes are remarkably high for both the thermal and non-thermal components, with the synchrotron component typically producing the higher flux after the first day of the outburst. 
The thermal emission mechanism from the photoionized ejecta \citep{2015ApJ...803...76C} 
usually dominates the radio light curves of novae with main-sequence companions \citep{2021ApJS..257...49C}. The ejecta in \tcb is expected to have low mass ($10^{-7}-10^{-6} M_{\odot}$; Paper I) and be rapidly expanding \citep{1992ApJ...393..289S}, so their ejecta should have low density and relatively low luminosity as a thermal radio emitter. 

The unabsorbed flux decreases quite rapidly. As shown in Fig.~\ref{tcb:fig-synchrotron-flux}, the unabsorbed thermal and nonthermal fluxes decline to the $\sim 1\un{Jy}$ level, characteristic of the quiescent state in our models (Table~\ref{tcb:table-quietfluxes-HDmodels}), within one day, except for the synchrotron from the RUN10 model (red solid line), characterized by high explosion energy. Its flux reaches the same level in about 10 days. 

\begin{figure}
  \centering 
  \includegraphics[width=\columnwidth]{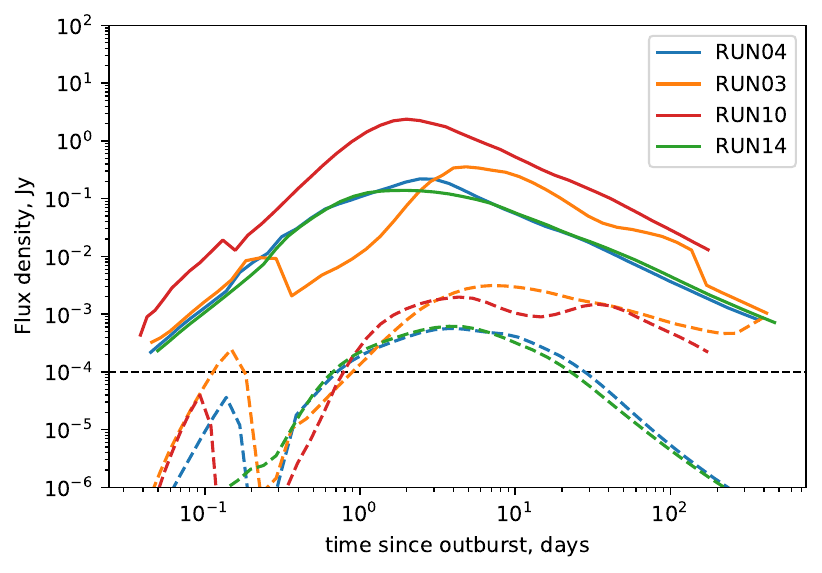} 
  \caption{%
       The same as in Fig.~\ref{tcb:fig-synchrotron-flux} with the Razin effect and absorption included.
       The dashed horizontal line shows approximately the flux density at $5\un{GHz}$ during the active state in the year 2016 \citep{2019ApJ...884....8L} and at $45\un{GHz}$ in the year 2024 \citep{2025A&A...702A.276P}. 
  }
  \label{tcb:fig-synchrotron-flux-abs}
\end{figure}

\begin{figure}
  \centering 
  \includegraphics[width=\columnwidth]{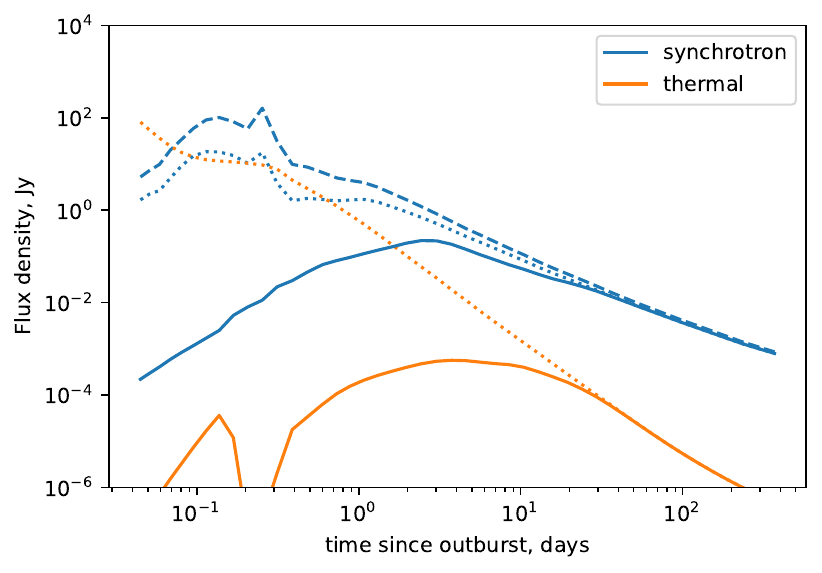} 
  \caption{%
       Comparison of different radio light curves for the RUN04 model at $1\un{GHz}$. Thermal (orange lines) and synchrotron (blue lines) contributions are shown as unabsorbed (dotted lines) and absorbed (solid lines) fluxes. For the synchrotron case, the dotted/solid line represents the unabsorbed/absorbed flux with the Razin effect, while the dashed line does not account neither for this effect nor for the absorption. 
  }
  \label{tcb:fig-freefree-synch-components}
\end{figure}
\begin{figure}
  \centering 
  \includegraphics[width=\columnwidth]{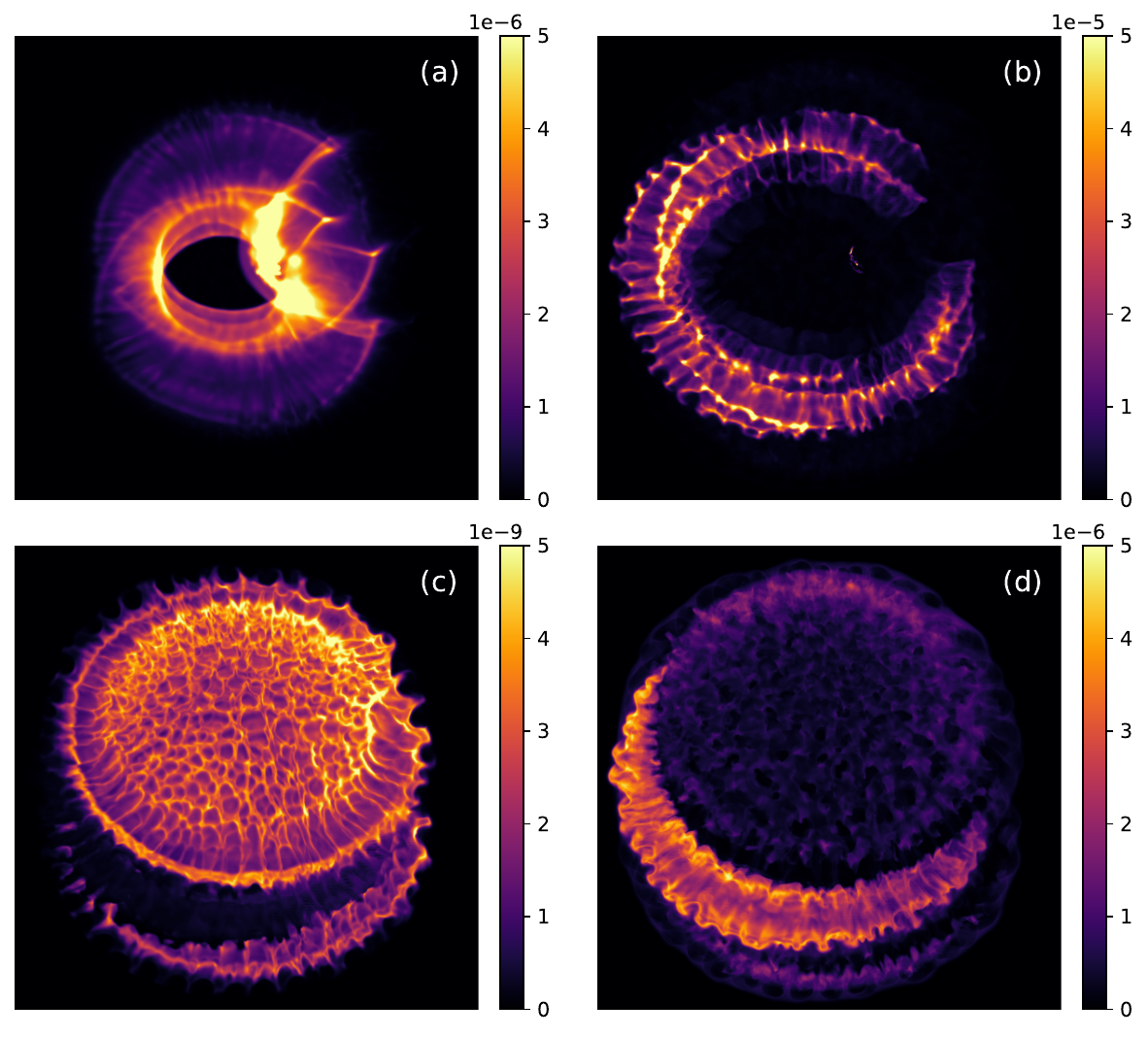} 
  \caption{%
       Radio maps at $1\un{GHz}$ for the RUN04 model on day 1.59 after the outburst. Images on the left (a, c) correspond to thermal emission, and those on the right (b, d) to synchrotron emission with the Razin effect. The top row shows the unabsorbed flux, while the bottom row accounts for the absorption. The color scale is in Jy/px. 
  }
  \label{tcb:fig-radio-img-with-absorption}
\end{figure}

The light curves with absorbed flux are shown in Fig.~\ref{tcb:fig-synchrotron-flux-abs}. They differ dramatically from the unabsorbed ones, and the cause is absorption. They grow initially and reach a maximum at about day 2 or 3 and then decrease similarly to the unabsorbed curves. As a result, the quiescent-state absorbed level ($\sim 10^{-4}\un{Jy}$, dashed black line in Fig.~\ref{tcb:fig-synchrotron-flux-abs} and Table~\ref{tcb:table-quietfluxes-HDmodels}) could be reached on a time scale of years. Remarkably, the thermal emission (dashed lines) is negligible during the first year in all our models. 

It is impressive how strongly the thermal radio waves are absorbed compared to the synchrotron radiation. This is clearly demonstrated by Fig.~\ref{tcb:fig-freefree-synch-components} as an example of the RUN04 model. This difference arises from the spatial distribution of the emitting and absorbing material, which affects the thermal and non-thermal components differently. Since synchrotron emission is produced predominantly near the forward shock, it is less affected by absorption and suffers smaller flux losses than thermal emission. 
Consequently, the observed radio emission is not simply a volume-integrated measure of the synchrotron luminosity: it preferentially samples those portions of the forward shock for which the line of sight has the smallest absorbing column. In fact, absorption suppresses the contribution from deeper layers, so that the observed flux is dominated by emission from regions closer to the observer, as clearly demonstrated by Fig.~\ref{tcb:fig-radio-img-with-absorption}.

In our models of \tcb, the Razin effect is prominent at early times. During the first week, it reduces the synchrotron intensity at $1\un{GHz}$ in the RUN04 model by approximately a factor of 3 (Fig.~\ref{tcb:fig-freefree-synch-components}; compare the blue dashed and dotted lines). We return to this effect later in the paper.

Absorption affects the radio flux from our models only at early times; the nova remnant becomes transparent to radio waves within about a month after the outburst (Fig.~\ref{tcb:fig-freefree-synch-components}). 
During this period, the absorbed morphology of our models differs significantly from the unabsorbed morphology (Figs.~\ref{tcb:fig-thermal-radio-maps} and \ref{tcb:fig-synchrotron-radio-maps}) in all our models. This is illustrated by Figs.~\ref{tcb:fig-radio-img-with-absorption} and \ref{tcb:fig-radio-maps-with-absorption} as also Animation~D4b, which show the combined free–free and synchrotron emission that remains after propagation of the radio waves through the absorbing medium at different epochs. 

\begin{figure}
  \centering 
  \includegraphics[width=0.697\columnwidth]{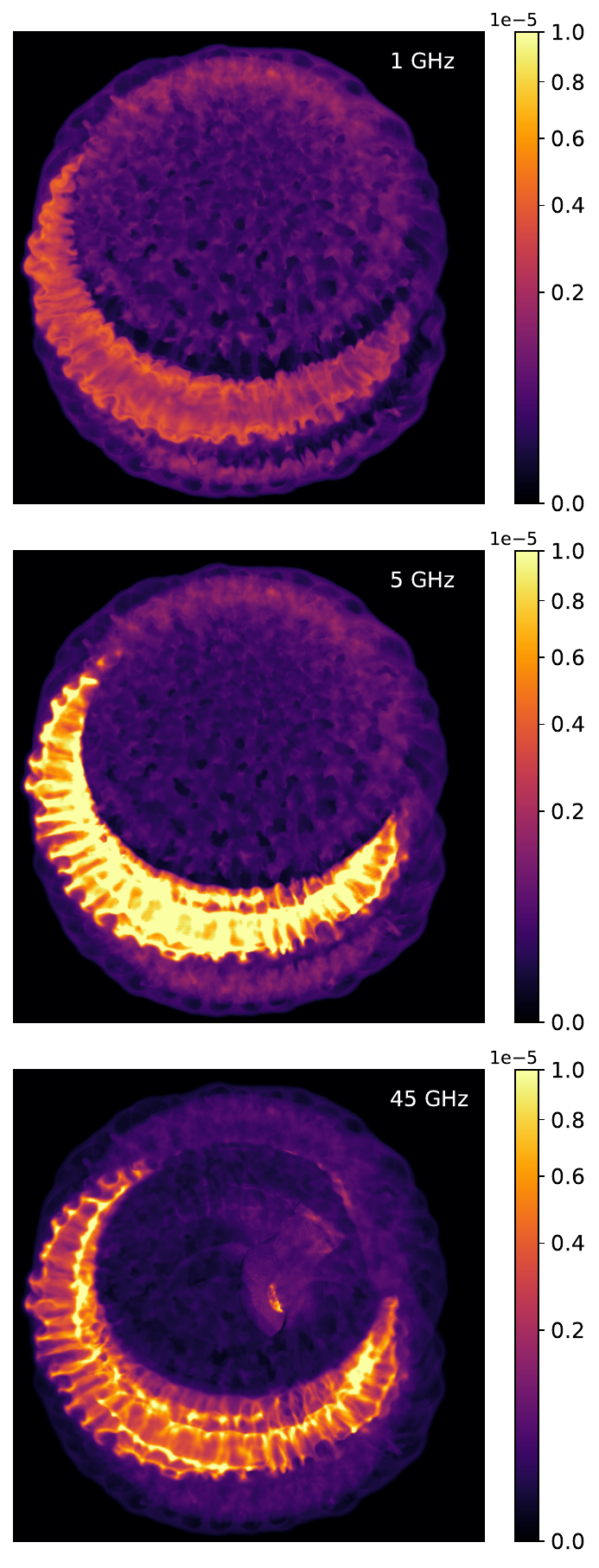} 
  \caption{%
        Total (free-free + synchrotron components) radio maps at $1\un{GHz}$ (top), $5\un{GHz}$ (middle), and $45\un{GHz}$ (bottom) for the RUN04 model on the day 1.59 after the outburst. Absorption and Razin effect are included. 
        The color scale is in Jy/px. 
  }
  \label{tcb:fig-radio-map-1-5-45-GHz}
\end{figure}
\begin{figure}
  \centering 
  \includegraphics[width=\columnwidth]{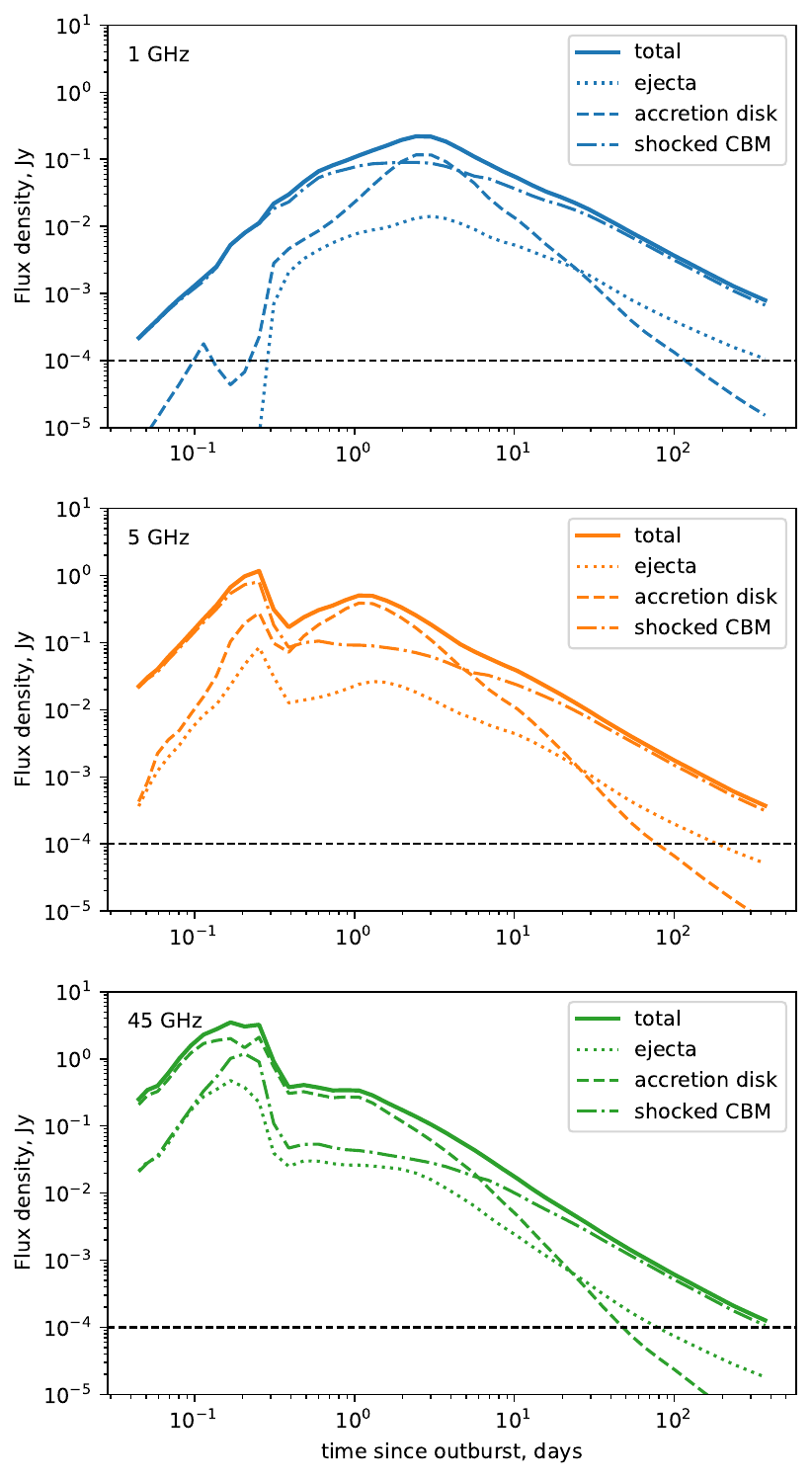} 
  \caption{%
       Total radio light curve (solid line) and its composition, namely, contributions from material of the ejecta, shocked accretion disk, and shocked CBM (consisting of the material from EDE and RG wind). RUN04 model, Razin effect and absorption included, i.e., what is expected to be observed. Frequencies: $1\un{GHz}$ (top), $5\un{GHz}$ (middle), $45\un{GHz}$ (bottom).
       Dashed horizontal line is the same as in Fig.~\ref{tcb:fig-synchrotron-flux-abs}.
  }
  \label{tcb:fig-flux-components}
\end{figure}

The bright ring visible on the synchrotron images in Fig.~\ref{tcb:fig-radio-img-with-absorption} is produced by emission from the material of the accretion disk that is swept away by the shock. This is clearly visible by comparison with the images for the model RUN14 which does not contain the disk: these images do not show such a bright ring (Fig.~\ref{tcb:fig-radio-maps-with-absorption}). One can notice on Animations~D4 how the shock displaces the material of the accretion disk.

In radio observations, a nova remnant could be resolved by VLBA at GHz frequencies in weeks or even days after the outburst \citep[e.g.][]{2006Natur.442..279O,2026arXiv260315480M}, when its angular size is $\gtrsim 10\un{mas}$. This corresponds to a physical size of about $10\un{AU}$ at a distance to \tcb.\footnote{Old superremnants of previous eruptions of \tcb and RS~Ophiuchi have sizes about $30$ and $70\un{pc}$ respectively \citep{2024ApJ...977L..48S,2025AJ....170...56S}.} The forward shock reaches this distance around day 4 (Fig.~\ref{tcb:fig-shock-front-params}). 
Absorption could still be effective (Fig.~\ref{tcb:fig-freefree-synch-components}), and interpretation of the radio images at this time should account for its spatial variation. As an illustration, the thin, extended visible part of a ring (Fig.~\ref{tcb:fig-radio-img-with-absorption}d) exhibits a pronounced SE–NW asymmetry and may appear jet-like solely because emission from other regions of the ring is absorbed (cf. Fig.~\ref{tcb:fig-radio-img-with-absorption}b). 
Therefore, an apparently bipolar or jet-like morphology in early radio images should not by itself be interpreted as evidence for intrinsically bipolar ejecta or jet launching. 
Morphological features due to absorption may be distinguished from intrinsic structures in a remnant by comparing images at different epochs and bands.

The evolution of visible morphology in our models with the accretion disk (Fig.~\ref{tcb:fig-radio-maps-with-absorption}) might be similar to RS~Oph \citep[e.g.][]{2006Natur.442..279O,2026arXiv260315480M}. Namely, it could appear with a SE-NW asymmetry like in a RUN04 model\footnote{The azimuthal directions correspond to our images where North is at the top and we arbitrarily chose the location of RG in NW part.} with a brighter SE part, then the almost full ring-like structure could emerge (corresponding to the emission in the accretion disk), after which the remnant becomes a nearly spherical diffuse nebula. If the density in the EDE and disk is higher, like in the RUN03 model, the remnant might also brighten along the North-South direction (which corresponds to directions perpendicular to the equatorial plane). 

\begin{figure*}
  \centering 
  \includegraphics[trim=5 70 20 5,clip,width=\textwidth]{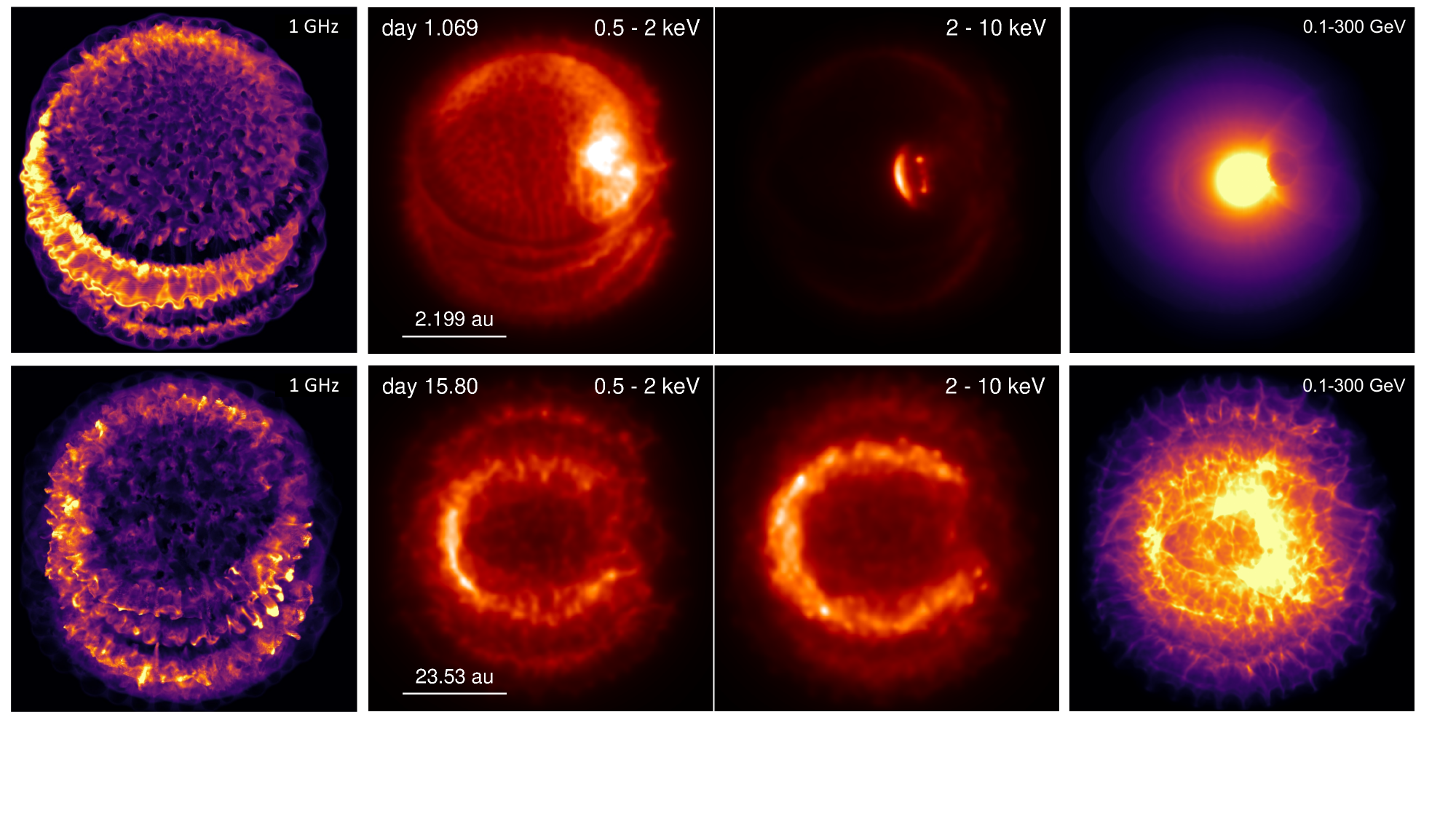} 
  \caption{%
       Images of \tcb on day 1 (upper row) and day 16 (lower row). Total radio (thermal + nonthermal with Razin effect; left column), soft and hard thermal X-rays (two middle columns), total \g-ray (leptonic + hadronic; right column). RUN04 model with absorption. In each frame, brightness is normalized to a value near the maximum; colors are scaled linearly.
  }
  \label{tcb:fig-multi-band-images}
\end{figure*}

Generally, the morphology varies with frequency because absorption depends on $\nu$, and the opacity varies across different regions of the nova remnant. Fig.~\ref{tcb:fig-radio-map-1-5-45-GHz} shows that the only outer layers of a remnant which are closer to the observer are visible at 1 GHz. The shocked accretion disk is more brighter at 5 GHZ because more photons penetrates through it. At 45 GHz, the internal structures become apparent. The next figure uncover when a certain component of a remnant dominates the emission.

Indeed, Fig.~\ref{tcb:fig-flux-components} shows the decomposition of the observed (i.e., absorbed) light curve for the basic RUN04 model into contributions from the ejecta, the accretion disk, and CBM. 
At 1 GHz, for most of the time, the dominant emission comes from the shocked EDE and RG wind 
(dot-dashed line), through the nonthermal emission of relativistic electrons. The contribution from the accretion disk (dashed line) is important around day 3. 
The ejecta signal (dotted line) remains approximately an order of magnitude smaller throughout the evolution because it is predominantly thermal, whereas the observed flux is dominated by synchrotron emission. 

Due to lower absorption, the peak of the light curve happens at the earlier times at higher frequencies (the same Fig.~\ref{tcb:fig-flux-components}). Therefore, the early multi-frequency observations could put constraints on the properties of the absorbing material. The visibility of accretion disk at $\nu=5\un{GHz}$ is from about the second half of the first day till the day 5. At 45 GHz, the accretion disk dominates emission from the beginning, in the model RUN04. From the end of the first week, the emission from the shocked CBM dominates in the total flux at both 5 and 45 GHz. 

It is worthwhile to compare the composition of the radio light curve in the RUN04 model with those of the X-ray and \g-ray emission. 
In soft X-rays, the ejecta is the major contributor to the flux throughout the first months as it may be seen from Fig.~7 in Paper~I and Animation~D2. In the hard band, the ejecta is important only from the end of the first day. Before then, as the same figure shows, the accretion disk dominates the hard X-rays. The shocked CBM (EDE and WD wind) emerges in the flux after about two/three months, in hard/soft photon energy ranges. 
Hadronic \g-rays are also dominated by the ejecta (Fig.~5 in Paper II and Animation~D3), while from the end of the first day until the end of the first week the leptonic \g-ray component becomes dominant. These photons are produced predominantly in the inner region of the nova remnant (i.e. mostly in the ejecta), since the inverse-Compton process occurs on nova photons whose density is higher close to the WD (see Animation~D3). After a week or so, the hadronic \g-rays reappear, with most of the emission generated in the shocked CBM.

As we can see, multi-frequency radio and multi-band observations at different epochs are important because different components and regions of the nova remnant are highlighted at different bands and times. This is further illustrated in Fig.~\ref{tcb:fig-multi-band-images}, which represents the remnant morphology in different wavelengths at two time moments. More detailed evolution in radio, X-rays, and \g-rays may be seen from animations in the Appendix~\ref{tcb-radio-app-movies}, which are synchronized and can be directly compared at each time moment. Of course, at present, sub-arcsecond structures cannot be spatially resolved in \g-rays, in contrast to the radio and X-ray domains. 
Taken together, these results illustrate a form of multi-wavelength tomography: radio synchrotron emission primarily traces the forward shock and shocked CBM, soft X-rays trace the dense shocked ejecta, while \g-rays probe different particle populations and interaction sites as the remnant evolves.

\begin{figure}
  \centering 
  \includegraphics[width=\columnwidth]{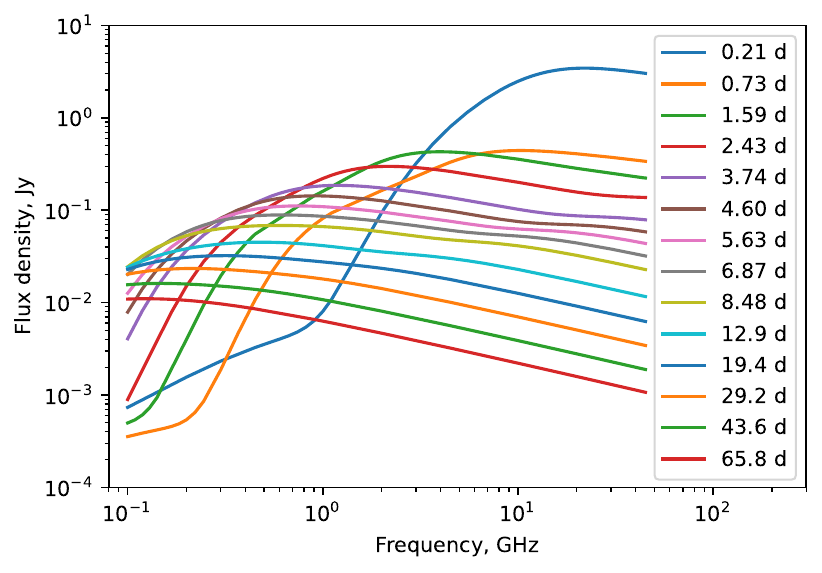} 
  \caption{%
       Evolution of the radio spectrum for RUN04 model of \tcb, Razin effect and absorption are included. Colors refer to time since the outburst, in days. 
  }
  \label{tcb:fig-spectra}
\end{figure}

Now, we return to radio emission. Fig.~\ref{tcb:fig-spectra} presents the evolution of the spectra between 0.1 and 10 GHz. As shown in Fig.~\ref{tcb:fig-synchrotron-flux-abs}, synchrotron emission dominates over thermal emission. The decrease in radio spectra at lower frequencies visible in Fig.~\ref{tcb:fig-spectra} is caused by Razin and absorption effects. 

The turnover frequency for the dominant thermal absorption is given by the equation~(\ref{tcorbor:eq-muff}): $\nu\rs{m}\propto nR^{1/2}T^{-3/4}$. This is the frequency at which the source experiences a transition from optically thick to optically thin regimes ($\tau=1$). Fig.~\ref{tcb:fig-spectra} shows that the characteristic frequency shifts toward lower frequencies with time, as expected, with the frequency decreasing by about an order of magnitude as the time increases by an order of magnitude. 
Indeed, for synchrotron emission at early times, it is roughly $\nu\rs{m}\propto R^{-1}$ for $n\sim \mathrm{const}$ (pre-shock density is dominated by EDE), $T\propto V^2$ (strong shock) and the shock speed $V\propto R$ (free expansion)\footnote{It is roughly $\nu\rs{m}\propto R^{-4}$ for the thermal emission of the ejecta (which is negligible in our models of \tcb), where $n\propto R^{-3}$ (constant ejected mass) and $T\propto n^{2/3}$ (adiabatic expansion).}
and $R$ scales linearly with time (Fig.~\ref{tcb:fig-shock-front-params}). 

\begin{figure}
  \centering 
  \includegraphics[width=\columnwidth]{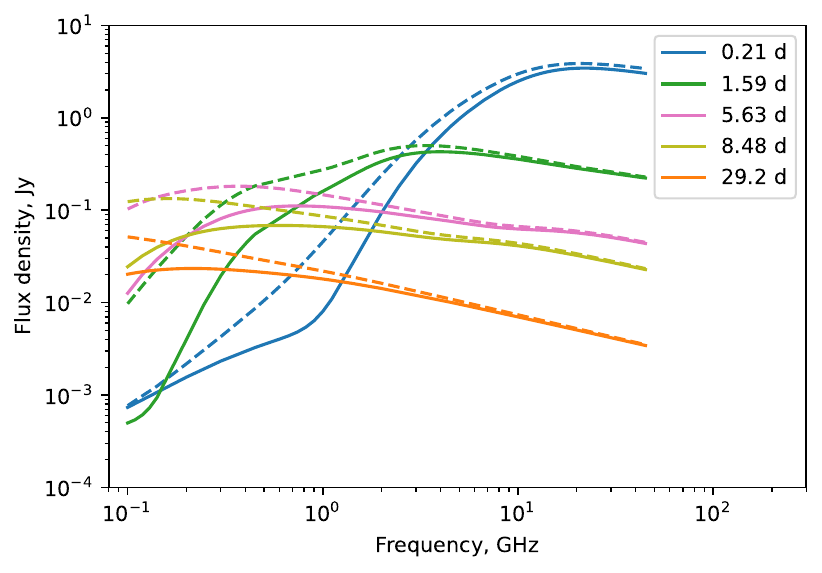} 
  \caption{%
       Influence of the Razin effect on absorbed radio spectra in the RUN04 model. 
       Solid lines are the same as in Fig.~\ref{tcb:fig-spectra}, with the Razin effect. Dashed lines do not account for this effect. 
  }
  \label{tcb:fig-spectra-vs-razin}
\end{figure}
\begin{figure}
  \centering 
  \includegraphics[width=0.98\columnwidth]{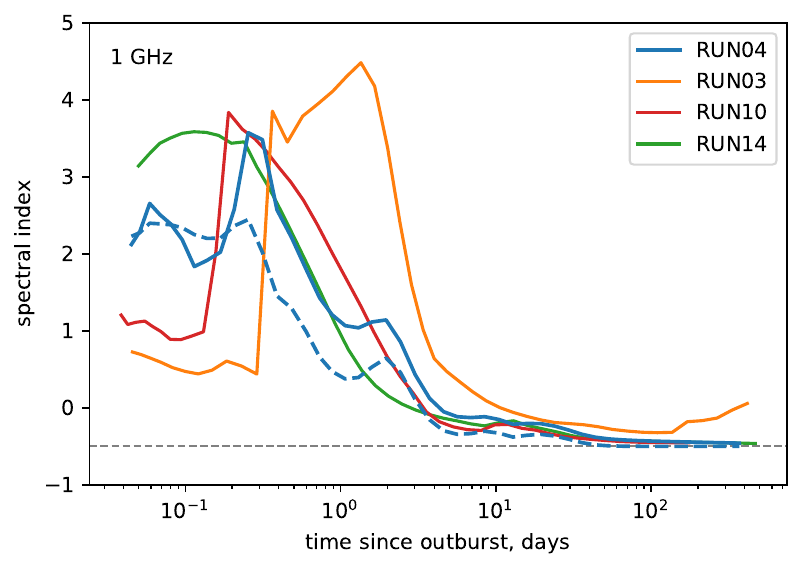} \\
  \includegraphics[width=0.98\columnwidth]{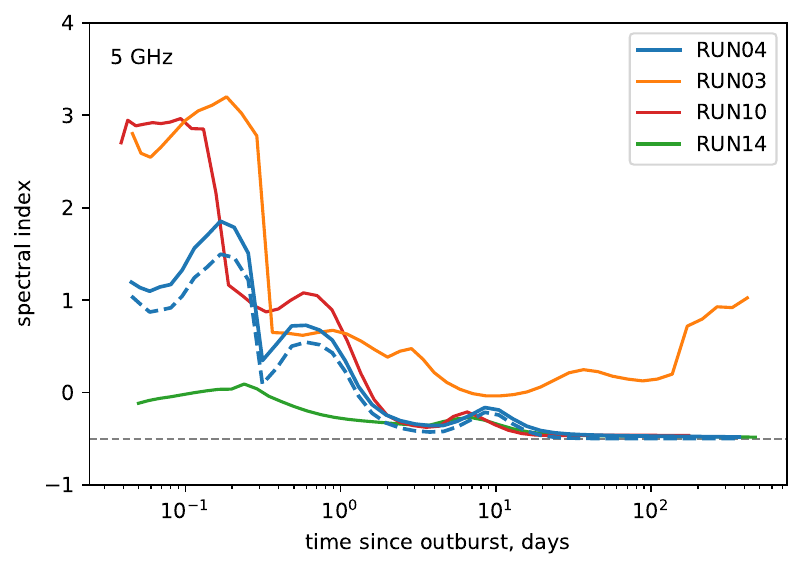} 
  \caption{%
       Evolution of the radio spectral index $\alpha$ (defined as $S\propto\nu^{\alpha}$) in our models of \tcb, measured from absorbed fluxes. Solid/dashed lines correspond to spectra with/without the Razin effect.
       Top: around $1\un{GHz}$ (between $0.9\un{GHz}$ and $1.1\un{GHz}$). Bottom: around $5\un{GHz}$ (between $4.5\un{GHz}$ and $5.5\un{GHz}$). Dashed horizontal line marks $\alpha=-0.5$.
  }
  \label{tcb:fig-spectral-index}
\end{figure}

As a consequence of such temporal changes in the spectra, the radio emission should peak at different frequencies at different times. Fig.~\ref{tcb:fig-spectra} shows, this effect may be visible in the light curves for $\nu\ll 10\un{GHz}$ where emission is heavily affected by the Razin effect and absorption. 

The slope of the orange line (day 0.73) in Fig.~\ref{tcb:fig-spectra} at frequencies $1$-$6\un{GHz}$ and green line (day 1.59) at $0.4$-$2\un{GHz}$ is about $\alpha\approx 0.8$ and $1.2$ respectively (for $\nu^{\alpha}$). Similar radio spectral indices were observed in the \tcb system in 2016 and 2024 \citep{2019ApJ...884....8L,2025A&A...702A.276P}. However, in the quiescent state, the emission is entirely thermal, whereas during nova evolution, synchrotron emission completely dominates.

It is worth noting that the Razin effect is a secondary factor contributing to the decrease in flux at long wavelengths; the dominant reduction is due to absorption (Fig.~\ref{tcb:fig-spectra-vs-razin}). Our \tcb models produce multi-zone emission, resulting in a low-frequency spectral shape that does not resemble a simple exponential cutoff, despite the locally exponential behavior of the emission from individual numerical cells. The radio spectra synthesized with the Razin effect have a higher radio index because of the increased steepness of the cut-off (Fig.~\ref{tcb:fig-spectra-vs-razin} and Fig~\ref{tcb:fig-spectral-index}).  The different physical dependence of the two processes (Razin effect and free-free absorption) makes multi-frequency monitoring particularly valuable: free-free absorption probes the emission measure and line-of-sight column, whereas the Razin effect is sensitive to the ratio $n\rs{e}/B$. Their combined evolution can therefore provide complementary constraints on the density and magnetic-field structure.

Fig~\ref{tcb:fig-spectral-index} shows the temporal variation of the radio spectral index around 1 and 5 GHz for our models of \tcb.  With time, $\alpha$ decreases from rather high values at the first hours and tends toward the classical $\alpha=-0.5$ value (corresponding to the assumed particle momentum slope $s=2$), reaching it finally in about a month. 
At both frequencies, the steepening of the spectral index up to approximately day 30 is primarily due to a reduction in absorption, while the Razin effect also affects the index, though to a lower extent.
At both frequencies, the radio index in our models becomes negative in few days after the nova outburst. Except of the model RUN03 (which has the EDE density in 100 times higher than in RUN04). In this model, the thermal radio emission is more important (less absorbed at higher frequencies); this is a reason why the yellow solid line at 5 GHz is close to $\alpha=0$ during few months. 
Early radio spectra of symbiotic recurrent nova V3890 Sgr \citep{2023MNRAS.523.1661N} also showed partially absorbed synchrotron emission with positive initial spectral indices evolving toward the optically thin synchrotron value $-0.5$, qualitatively consistent with our predictions for \tcb. 
In V745 Sco, instead, the radio spectral index decreases from about unity during the first few days to about zero by the end of the second week and then remains approximately constant \citep{2024MNRAS.534.1227M}, suggesting a thermal origin for the radio emission, as in our model RUN03. In RS~Oph, the index $\alpha$ is positive up to day 10 and, from day 24, also stays around zero \citep[Fig.~3 in][]{2023MNRAS.523..132D}.

With time, the thermal component of the radio flux could affect the radio spectral index in \tcb. For example, after about 100 days, $\alpha$ in the RUN03 model (orange line in Fig~\ref{tcb:fig-spectral-index}) increases again. This happens because the thermal component of the radio emission becomes more efficient at this epoch (Fig.~\ref{tcb:fig-synchrotron-flux-abs} dashed line).

\section{Conclusions}
\label{tcb-radio:sect-concl}

In this paper, we synthesized the expected radio emission from the highly anticipated outburst of the recurrent nova T~Coronae Borealis, utilizing 3D hydrodynamic models coupled with multi-zone particle acceleration. Our models track the complex interplay between thermal and non-thermal emission and absorption as the forward shock interacts with the structured circumbinary medium and accretion disk. We highlight the following key predictions for upcoming radio observation campaigns. 

\begin{itemize}
\item	Our synthesis of the pre-outburst quiescent state reveals that while the accretion disk generates more than 99\% of the unabsorbed radio emission, this radiation is almost completely absorbed within the disk itself. The observable quiescent flux instead originates primarily from the outer layers of the red giant wind and the equatorial disk enhancement, yielding absorbed flux densities ($0.03-0.3\un{mJy}$ at 45 GHz) that are consistent with recent pre-outburst measurements. Furthermore, we find that the absorbing role of the unshocked CBM drops rapidly after the outburst, becoming negligible once the forward shock traverses the accretion disk within the first six hours.
\item	During the first days following the eruption, the intrinsic radio emission is characterized by extreme unabsorbed radio fluxes (reaching $\sim10^2-10^3\un{Jy}$ at GHz frequencies). However, this emission is heavily obscured by absorption from the shocked accretion disk, the equatorial disk enhancement, and the dense ejecta. 
\item	An important result enabled by the 3D nature of our simulations is that thermal and synchrotron emission are absorbed very differently. Thermal emission is much more strongly absorbed than synchrotron emission. This occurs because the emitting and absorbing materials have different spatial distributions. Synchrotron emission is produced mainly in the outer layers near the forward shock and is less affected by absorption. In contrast, thermal emission comes from deeper layers and is strongly suppressed by the material in front of it.
\item	In addition to free-free absorption and synchrotron self-absorption, the Razin-Tsytovich effect further suppresses low-frequency synchrotron emission during the initial expansion. This effect reduces the 1 GHz flux by approximately a factor of 3 in our reference model during the first week and steepens the low-frequency cutoff. The effect could be neglected at frequencies above 5 GHz, even during the earliest phase. With time, the Razin frequency decreases.
\item	The peak of the radio light curve at 1 GHz is predicted to occur around days 2–5 in different models. At frequencies below 1 GHz, the light curves are expected to peak at later times because they are strongly affected by absorption.
\item The absorption is lower at higher frequencies. Therefore, the radio flux reaches its peak at 5 GHz and 45 GHz earlier. At higher $\nu$, the internal structures become visible earlier comparing to lower $\nu$. 
\item Material of different components of the nova remnant (shocked EDE and RG wind, accretion disk, ejecta) dominates the flux during different periods. For example, the accretion disk dominates over the shocked CBM during one day at 1 GHz and during almost a week at 5 or 45 GHz.
\item	The nova remnant is predicted to become transparent to GHz radio waves within approximately one month at 1 GHz. We should expect the radio spectral index to evolve from initially positive values, driven by opacity and the Razin effect, down to the classic optically thin synchrotron value of $-0.5$ after several weeks. 
\item	Contrary to novae with main-sequence companions, the observed radio flux in \tcb will remain non-thermal during the first year. The thermal component could become observationally significant only at later times in models featuring an exceptionally dense EDE. 
\item 3D treatment uncovers that, the observed radio morphology of \tcb can be qualitatively different from the intrinsic morphology at early times, due to absorption. Absorption can make only a portion of the ring visible at frequencies $\lesssim 5\un{GHz}$ and therefore produce an apparent SE–NW asymmetry or even mimic a jet-like structure. 
\item The morphology evolution varies in detail across models, but an evolutionary path could follow this sequence: an incomplete ellipse marking the plane of the accretion disk -- a complete ellipse or elliptical disk with asymmetry -- brightening in the direction perpendicular to the orbital plane -- diffuse spherical nebula.
\item	We see from our simulations that polarized emission is generally not detected in novae because cosmic-ray acceleration generates strong, disordered magnetic fields that depolarize the synchrotron emission. The ordered magnetic field of the red giant companion -- if strong enough to dominate the cosmic-ray-generated field in its vicinity -- may induce a brief window for detecting polarized radio emission during the first few days after eruption. 
Early multi-frequency polarimetric monitoring could provide valuable constraints on the magnetic-field strength and topology in the red-giant wind and on the relative contribution of ordered and shock-amplified magnetic fields.
\item Our synthetic radio data highlights a strong spatial and temporal complementarity with X-ray and \g-ray emission. While soft X-rays and hadronic \g-rays are primarily generated within the ejecta, the radio band effectively isolates ongoing interaction of the forward shock with the accretion disk and outer CBM. Simultaneous multi-band monitoring will therefore be crucial for reconstruction of the \tcb environment.
\end{itemize}

Radio images synthesized from 3D HD data reveal prominent differences between the intrinsic (unabsorbed) and observable (absorbed) structures of the nova remnant during the first weeks following the outburst. While intrinsic synchrotron emission arises primarily from a thin layer at the forward shock, and thermal emission originates from deeper, denser regions, strong line-of-sight absorption severely alters the observed shape. Because of this absorption, the early visible morphology may exhibit a pronounced asymmetry or appear jet-like, solely because emission from other parts of the expanding ring is obscured. As the remnant expands and becomes transparent, the visible morphology is predicted to evolve from this initial asymmetric shape into a fuller ring-like structure reflecting the accretion disk and EDE, and eventually into a nearly spherical diffuse nebula or two-bubble structure in case of high density in the equatorial plane, similar to the morphological evolution seen in RS~Ophiuchi.

To facilitate a comprehensive view of the outburst, we provide multi-band animations that allows direct comparison of the \tcb models across different parts of the electromagnetic spectrum. These supplemental materials detail the temporal evolution of the nova remnant simultaneously in radio, X-rays, and \g-rays, highlighting how different regions and physical components of the expanding shock dominate at various wavelengths and epochs. 

The radio luminosity remains sensitive to assumptions concerning the electron-to-proton ratio, acceleration efficiency, magnetic-field amplification, and transport of relativistic electrons. The present calculations therefore provide predictions for a fiducial acceleration model rather than a unique determination of these parameters. Future radio observations, combined with X-ray and gamma-ray constraints and spatially resolved spectroscopy, can break these degeneracies and provide a quantitative test of particle acceleration in the T CrB shock.

\begin{acknowledgements}
O.P. acknowledges support from INAF 2023 RSN4 Theory grant and its extension by INAF. 
L.C. is grateful for support from NSF grant AST-2107070 and NASA grants 80NSSC23K0497 and 80NSSC25K7334.
F.B. acknowledges partial support from INAF Grant AF2024.
O.P. and T.K. thank the Armed Forces of Ukraine for providing security to perform this work. 
V.B. serves in these Forces; he was involved in the initial studies and analysis of methods for synthesizing radio emission. 
This research used the HPC system MEUSA at the SCAN (Sistema di Calcolo per l’Astrofisica Numerica) facility for HPC at INAF-Osservatorio Astronomico di Palermo, Italy.
\end{acknowledgements}

\bibliographystyle{aa}
\bibliography{tcorbor}

\begin{appendix}  

\section{On the energy spectrum of cosmic rays}
\label{tcb-radio-app-e-spectrum}

We follow the approach described in Sect.~2 of Paper~II for the calculation of the energy distribution of radio-emitting electrons. Namely, in each cell over the shock surface we take the power law $N(E)dE=KE^{-s}dE$. 
The normalization $K$ in the electron energy spectrum is taken $K\rs{ep}$ times the normalization $K\rs{cr}$ of the proton energy spectrum. 
The normalization $K\rs{p}$ of the momentum distribution of protons $K\rs{p}p^{-s}dp$ is given by the condition that the integral of the proton spectrum over $p$ gives a fraction $\xi\rs{in}$ of the total number density of plasma at the shock $n_2$, i.e., $K\rs{p}\simeq \xi\rs{in} n_2 p\rs{in}$ where $\xi\rs{in}$ and $p\rs{in}$ are the injection efficiency and the injection momentum for protons. The normalization of the energy distribution is related to the normalization of the momentum distribution as $K\rs{cr}=K\rs{p}c^{s-1}$. Thus, $K=K\rs{ep}\xi\rs{in} n_2 p\rs{in} c^{s-1}$.

The sum of values of $K$ in each cell across the shock provides the overall normalization $K\rs{tot}$ at a given time. The value of $\xi\rs{in}$ is taken to provide a given value of the overall acceleration efficiency $\xi\rs{cr}$, which is a fraction of the shock kinetic energy transferred to CRs. 
In the present paper, the values are the same as in Paper II: $K\rs{ep}=0.01$, $s=2$, $\xi\rs{cr}=0.1$, constant in time. We note that these values are fiducial assumptions adopted from Paper II rather than quantities independently constrained by the calculations or observations. Change of $K\rs{ep}$ results in the same change of the synchrotron flux and $\mu\rs{ssa}$ because they are just proportional to $K\rs{ep}$. Slightly different value of $s$ does not affect much the results; after the outburst, it could be determined from the observed synchrotron spectra. By changing $\xi\rs{cr}$ we change both the normalization of the particle spectrum and the maximum energy ($E\rs{max}$ depends on $\delta B$ which depends on $\xi\rs{cr}$). Our HD simulations does not account an eventual feedback of CRs, therefore, we cannot consider values of $\xi\rs{cr}$ considerably higher than that. Further discussion is present in the Appendix~B of Paper II.

The normalization $K\rs{tot}$ varies in time. At each time moment, we consider it spatially uniform.  This is due to our assumption that diffusion of accelerated electrons through the shocked volume is much faster than their advection downstream of the forward shock. Indeed, the propagation length $\ell$ of the GeV-TeV electrons is considerably larger than the radius $R$ of a nova remnant. Let us take $\ell\simeq \sqrt{2\,D\,\Delta t}$ and adopt, for particle propagation, the diffusion coefficient which is nearly the Galactic one $D_1=2\E{28}\,(EB\rs{ref}/B E\rs{ref})^{0.5}\un{cm^{2}/s}$ with $E\rs{ref}=1\un{GeV}$ and $B\rs{ref}=3\un{\mu G}$. Then $\ell\simeq 920\,t\rs{d}^{0.5}(E\rs{GeV}/B\rs{mG})^{0.25}\un{AU}$ where $t\rs{d}$ time in days, $E\rs{GeV}=E/1\,\mathrm{GeV}$, $B\rs{mG}=B/1\,\mathrm{mG}$. If $B\sim1\un{G}$ (as it could be at the early stage, Figs.~\ref{tcb:fig-shock-front-deltaB} and \ref{tcb:fig-synchrotron-MFordered}) then $\ell\simeq 160\,t\rs{d}^{0.5}\un{AU}$. The radius $R$ reaches $160\un{AU}$ in about 100 days (Fig.~\ref{tcb:fig-shock-front-params}); by then, however, MF is of the order of mG (Fig.~\ref{tcb:fig-shock-front-deltaB}) and $\ell\simeq 920\un{AU}$ in one day.

\begin{figure}
  \centering 
  \includegraphics[width=\columnwidth]{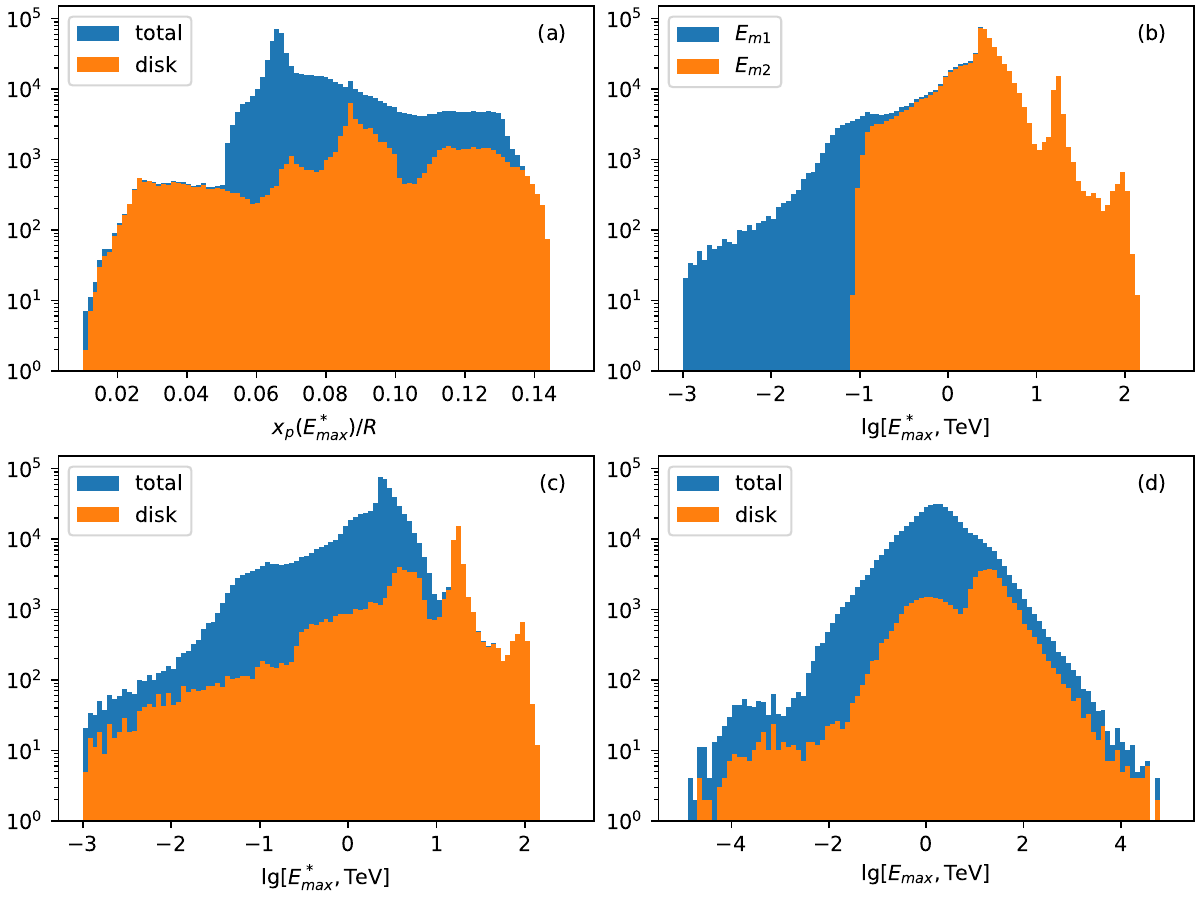} 
  \caption{%
       Histograms showing properties of CRs with highest energies for the RUN04 model of \tcb at 5 hours (0.21 day) after the nova outburst. In subplots a, c, and d, the local values for each numerical cell across the forward shock are shown in blue color; the orange color marks the cells interacting with the accretion disk. 
       {\bf (a).} Ratio of the acceleration length scale to the shock radius $x\rs{p}(E\rs{max}^*)/R$; $E\rs{max}^*$ and $R$ are local values for each numerical cell. 
       {\bf (b).} Maximum energy determined in each cell by the limited acceleration time ($E\rs{m1}$, blue color) or time-to-grow of the Bell modes ($E\rs{m2}$, yellow color).
       {\bf (c).} Maximum energy $E\rs{max}^*$ determined under the assumption of uniform flow upstream and downstream.
       {\bf (d).} Maximum energy $E\rs{max}$ estimated for each cell by accounting for the gradient of the flow speed downstream (see Paper II for details).
  }
  \label{tcb:fig-xp-emax}
\end{figure}

Note that for the particle acceleration, we consider the Bohm diffusion. 
Therefore, the acceleration length scale $x\rs{p}\simeq D/u$ is much shorter. Indeed, with the Bohm diffusion coefficient $D=r\rs{L}c/3$, it is $x\rs{p}\simeq 4.5\E{-3}E\rs{GeV}/(V_{5000}\,B\rs{mG})\un{AU}$ where $V_{5000}=V/5000\un{km\,s^{-1}}$. It is small, $x\rs{p}\simeq 0.045\un{AU}$, even for CRs with energy $E\simeq 10\un{TeV}$ in $B\simeq1\un{G}$ around the shock with $V\simeq 5000\un{km/s}$, i.e. $x\rs{p}\ll R$ even for high $E$ and even very early after a nova outburst. Indeed, Fig.~\ref{tcb:fig-xp-emax}a shows a histogram for $x\rs{p}(E\rs{max}^*)/R$ for each numerical cell across the forward shock. As we see, the acceleration length scale for the highest energy CRs is within the range $\approx 2$-$14\%$ of the shock radius even at a very early stage. 

At this point, it is worth considering other properties of the highest-energy CRs. As shown in Paper II, CRs could reach PeV energies during the first hours after the nova outburst. The maximum energy $E\rs{max}^*$ is calculated in Paper II as the minimum of values given by three processes, namely, the limited acceleration time ($E\rs{m1}$), the time-to-grow for the Bell instability ($E\rs{m2}$), and radiative losses. The last process is not important for protons. Fig.~\ref{tcb:fig-xp-emax}b demonstrates that the second process (yellow color on the plot) is dominant (i.e. $\mathrm{min}[E\rs{m1},E\rs{m2}]=E\rs{m2}$) in the determination of the proton maximum energy. This means there is enough time for acceleration to maximum energies even during the first hours, in particular, due to high MF. Fig.~\ref{tcb:fig-xp-emax}c shows that $E\rs{max}^*$ calculated under a common approach to diffusive shock acceleration in the uniform flow upstream and downstream provides $E\rs{max}^*$ up to $100\un{TeV}$ at $t=5$ hours after the outburst; with the highest energies of $10$-$100$ TeV reached at the shock propagating through the dense accretion disk (yellow color). When the effect of the downstream gradient in the flow speed \citep{2024A&A...688A.108P,2026A&A...709A.175P} is taken into account, as in Paper II, the CR energy reaches the PeV scale in some cells (Fig.~\ref{tcb:fig-xp-emax}d). In \tcb, this occurs around the ejecta protrusions (Paper II), where this gradient is strongest. High gradients of the flow speed can also develop in shocks interacting with dense material, due to radiative losses \citep{2026A&A...709A.175P}. However, in our models of \tcb, the plasma temperature does not fall below $10^6\un{K}$ even when the shock propagates through the dense accretion disk. Therefore, the radiative scenario does not occur in \tcb.

\section{Remark on evolution of the MF disordered component downstream of a nova shock}
\label{tcb-radio:sect-app-dBevol}

Distribution of $\delta B$ downstream of the SNR shock, as given by the continuity equation for waves, is considered in detail by \citet[][Sects.~3 and 4]{2017MNRAS.470.1156P}, by employing nonlinear Landau damping and growth due to resonant interaction of Alfv\'en waves with accelerated protons. In this approach, the distribution of disordered MF downstream of the Sedov shock is roughly similar to the distribution of thermal pressure (figures~4 and 5 in this reference), i.e., the magnetic pressure $\delta B^2/8\pi$ does not vanish even in the deep interior of a remnant.
In an alternative approach, the distribution of $\delta B(r)$ from Eq.~(\ref{tcbsynch:dBevol}), which is determined by the evolution of HD structure, is similar to the density profile that rapidly drops behind a Sedov shock \citep[e.g. Fig.~2 in][]{2016MNRAS.456.2343P}. The downstream decay of CR-generated $\delta B$ is faster in the second approach than in the first. Therefore, we use (\ref{tcbsynch:dBevol}) as a more stringent condition that constrains the structure of the MF inside the nova remnant. 

This choice is also supported by a comparison of the relevant timescales $\tau=1/\Gamma$. The damping rate due to the Landau process is $\Gamma\rs{nl}\simeq 0.1 v\rs{A}k\rs{min}(\delta B/B)^2$ 
\citep[e.g.][eq.~33]{2017MNRAS.470.1156P} with the lowest wavenumber $k\rs{min}\sim 1/R$. The rate due to flow expansion is $\Gamma\rs{ad}\simeq u_2/3R=V/12R$ where we assumed the length-scale for MF adiabatic decay $\sim0.3R$ (for comparison, the length scale for the density decay is $\sim0.1R$). Therefore, 
\begin{equation}
 \frac{\tau\rs{ad}}{\tau\rs{nl}}
 \sim 0.1\,\frac{(\delta B/B)^2}{V_{3000}}\, \frac{\delta B\rs{mG}}{n^{1/2}}
\end{equation}
where $V_{3000}$ the shock speed in units $3000\un{km/s}$, $\delta B\rs{mG}$ the MF strength in mG. By taking $\delta B\sim 1\un{mG}$, $n\sim 10^3\un{cm^{-3}}$, $\delta B/B\sim 3$, we have $\tau\rs{ad}/\tau\rs{nl}\sim 0.03$. With time, the MF scales as $\delta B\propto n^{1/2}$, therefore the ratio remains similar during the evolution of our models. 
Thus, adiabatic losses are dominant: the flow moves the magnetic field away from the shock faster than kinetic damping can remove energy from it. 

\citet{2005ApJ...626L.101P} considered the limit of zero growth of waves downstream and derived the expression for the damping length scale $l\rs{d}$ (their equation 7): 
\begin{equation}
 l\rs{d}\sim 6.6\, \frac{V_{3000}\, n^{1/2} \,
 E\rs{10TeV}}{\delta B\rs{mG}^{2}}\, \un{AU}
 \end{equation}
where we substituted the resonant wavelength $\lambda$ by the gyro-radius for particles with maximum energy $r\rs{g}(E\rs{max})=2.2 E\rs{10TeV} B\rs{mG}^{-1}\un{a.u.}$, as also $\delta B(\lambda)\sim \delta B$ because the overall MF strength is dominated by the largest resonant wavelength. By using the same numbers for estimates as above and $E\rs{max}=10\un{TeV}$, we obtain $l\rs{d}\sim 210\un{a.u.}$, which is larger than the adiabatic decay length scale $\sim0.3R$. If, in addition, the grow rate is not zero, the damping length scale is even larger.

\section{Radio images}
\label{tcb-radio-app-images}

\begin{figure*}
  \centering 
  \includegraphics[width=\textwidth]{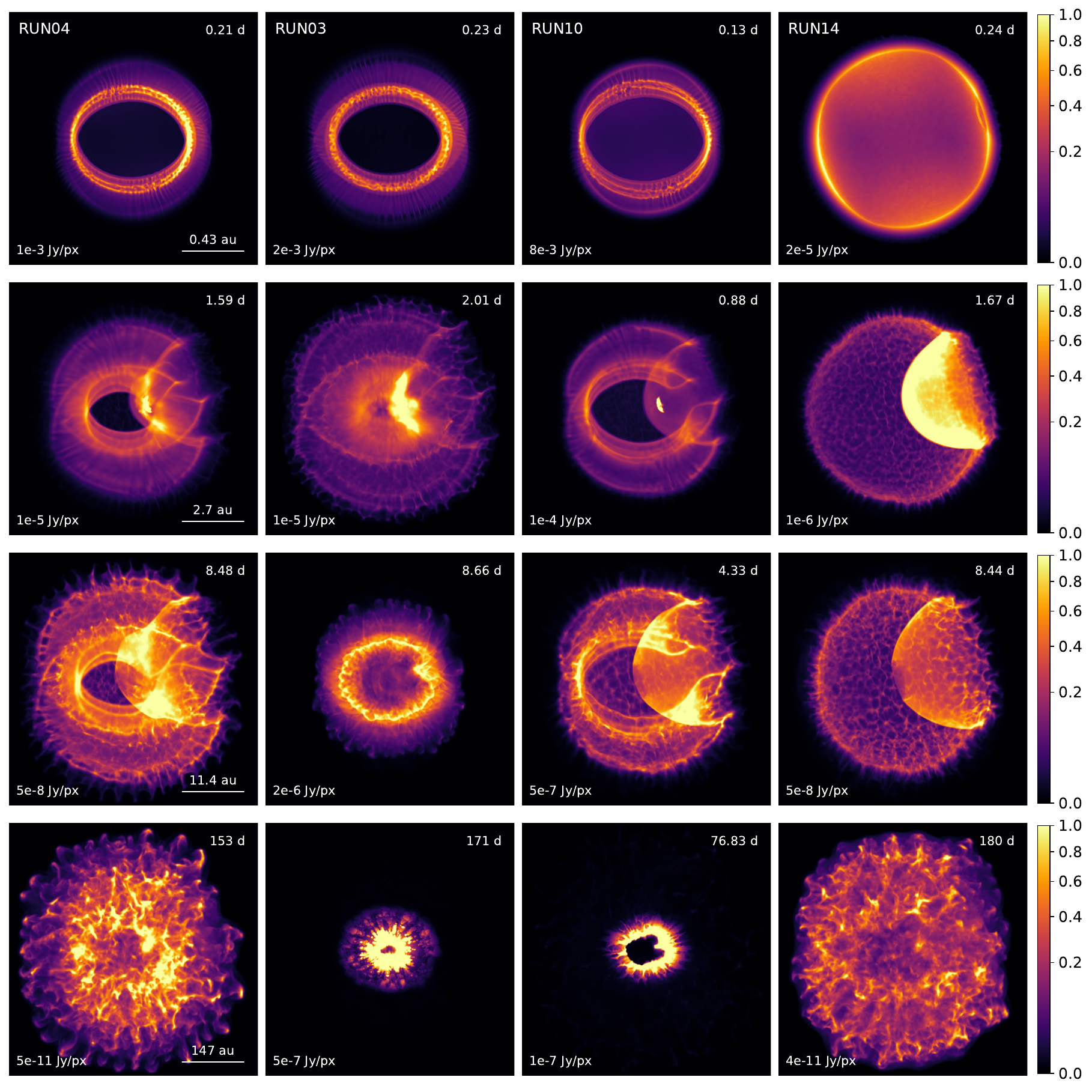} 
  \caption{%
       Images of different models of \tcb (columns) in thermal free-free emission at $1\un{GHz}$ for several time moments (rows). Absorption is not accounted for. The color scales are in normalized units, i.e., $S/S\rs{norm}$. The normalization value $S\rs{norm}$ is shown in the lower-left corner of each frame. The spatial scale is shown in the lower-right corner of the leftmost frame. It is the same for all frames in a row. 
  }
  \label{tcb:fig-thermal-radio-maps}
\end{figure*}
\begin{figure*}
  \centering 
  \includegraphics[width=\textwidth]{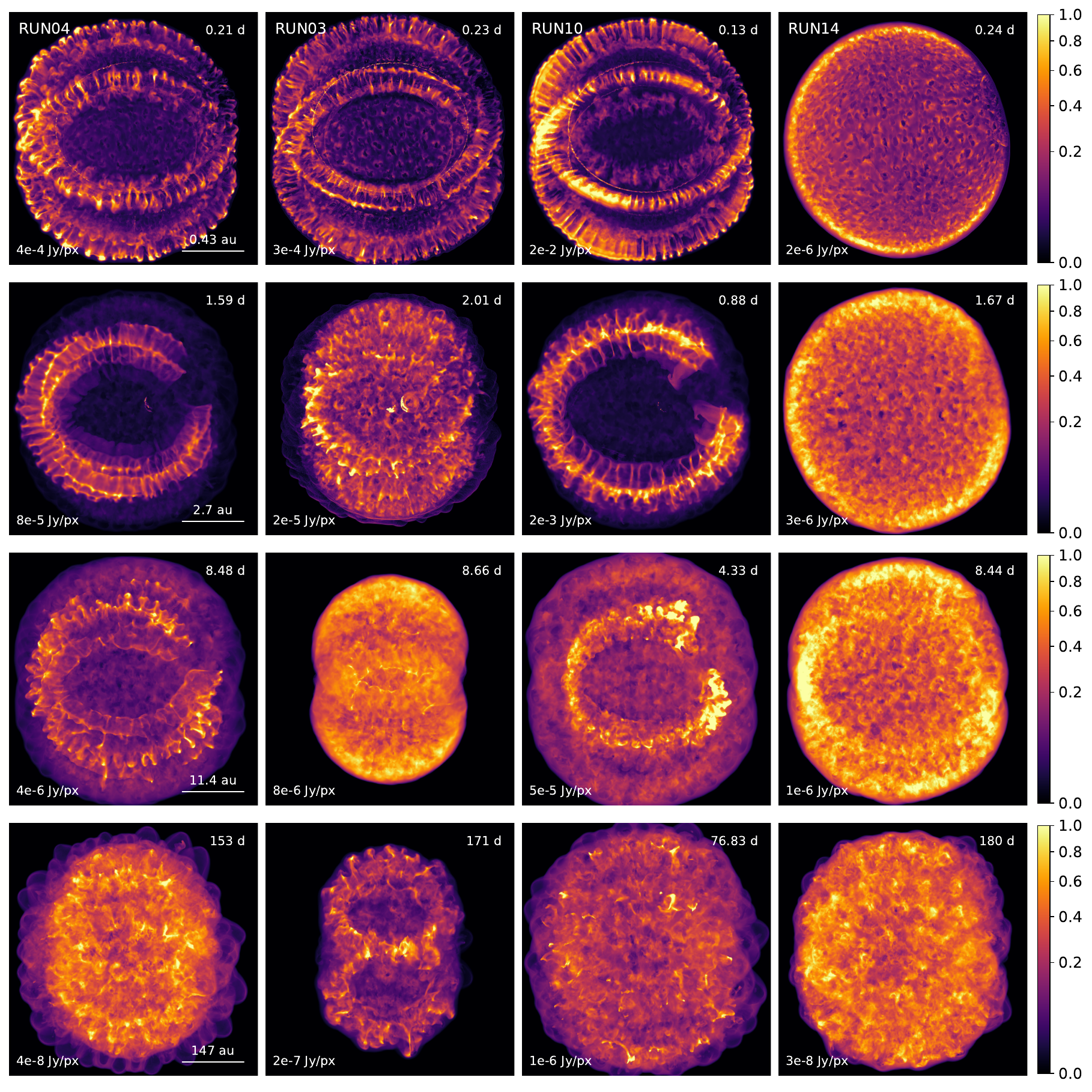} 
  \caption{%
       Maps of the Stokes parameter $I$ at $1\un{GHz}$ for different \tcb models (columns) at several time moments (rows). Absorption is not included, whereas the Razin effect is taken into account. 
       The color scales are in normalized units as in Fig.~\ref{tcb:fig-thermal-radio-maps}.
  }
  \label{tcb:fig-synchrotron-radio-maps}
\end{figure*}
\begin{figure*}
  \centering 
  \includegraphics[width=\textwidth]{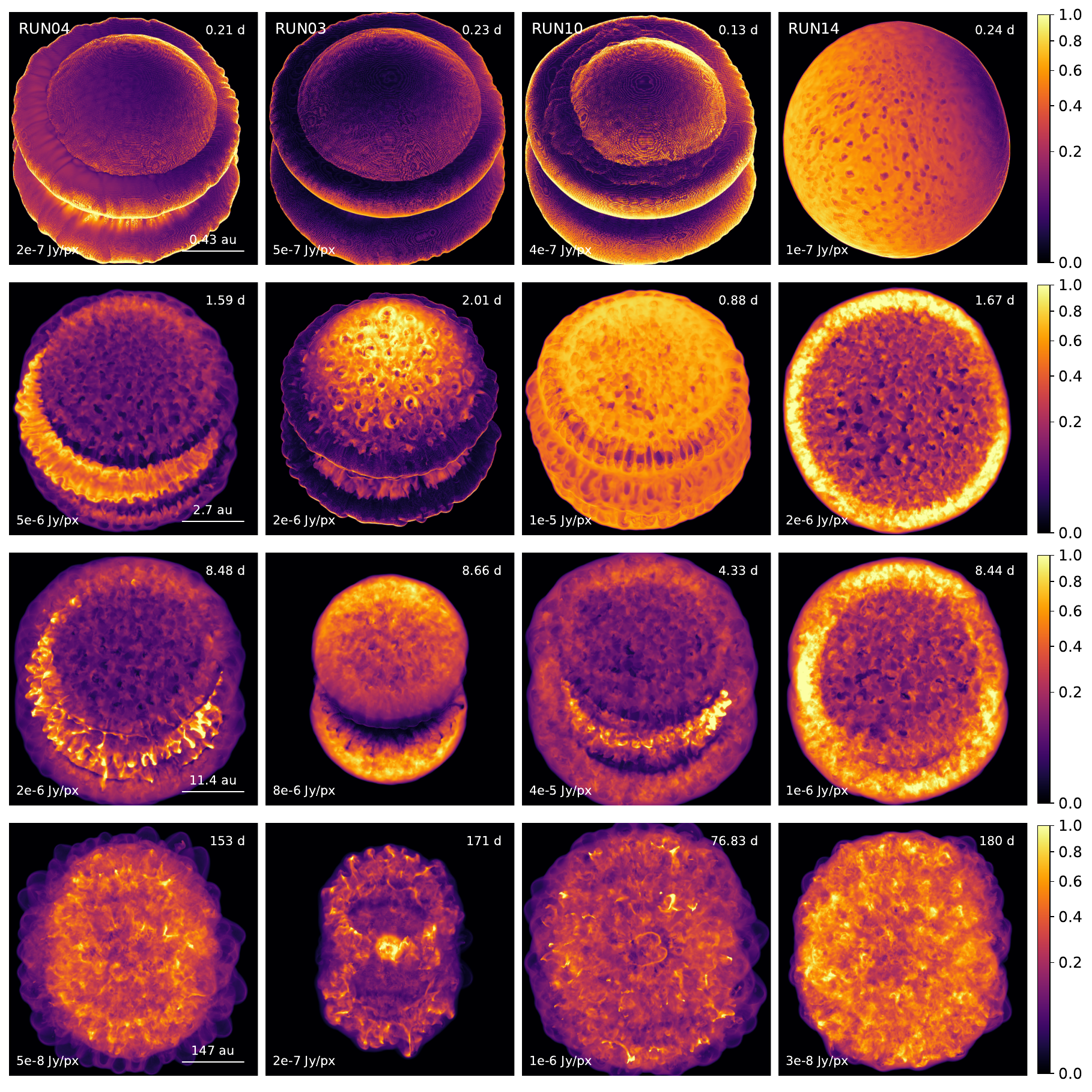} 
  \caption{%
       Radio maps (thermal free-free + nonthermal synchrotron) at $1\un{GHz}$ for different \tcb models (columns) for several time moments (rows). Models account for the Razin effect and absorption. 
       The color scales are in normalized units as in Fig.~\ref{tcb:fig-thermal-radio-maps}.
  }
  \label{tcb:fig-radio-maps-with-absorption}
\end{figure*}

Sequences of radio images from our models of \tcb. Fig.~\ref{tcb:fig-thermal-radio-maps}: unabsorbed thermal free-free component of emission, without absorption. Fig.~\ref{tcb:fig-synchrotron-radio-maps}: unabsorbed non-thermal synchrotron emission component with Razin effect included. 
Fig.~\ref{tcb:fig-radio-maps-with-absorption}: sum of the two components (thermal and non-thermal), absorbed. 
The images in the same row of each of Figs.~\ref{tcb:fig-thermal-radio-maps}-\ref{tcb:fig-radio-maps-with-absorption} have the same black square size, allowing the sizes of the emitting regions to be compared directly. 
As in Papers I and II, images are synthesized for a random orbital phase (corresponding to initial rotation of the system about $z$-axis by $20\degr$, so, RG is located in the NW part of each frame), for an orbital inclination $55\degr$ and a distance $890\un{pc}$. 

We would like to note that the \g-ray images in Figs.~6 and 8 in Paper~II are produced for the orbital inclination $35\degr$; all other results in Paper~II are presented for the inclination $55\degr$.

\section{Multi-band animations}
\label{tcb-radio-app-movies}

All animations are produced from the reference RUN04 model of \tcb. The color scale on each image is normalized to a value close to the maximum. The horizontal white line represents the physical length scale for the assumed distance to \tcb of 890 pc. At this distance an angular separation of 0.45" corresponds to a distance of 400 AU. Surface brightness images are shown for the orbital phase corresponding to an initial rotation of $20\degr$ about the $z$-axis and an orbital inclination of $55\degr$, obtained by a second rotation about the $x$-axis. In the Cartesian coordinate system, the $x$-axis points to the right, and the observer is located along the negative $y$-axis. In the simulation setup, RG is located in the positive $x$.

The associated movies are available online.

Animation~D1. Evolution of density in the reference model. 2D cross-section in the $x$-$z$ plane is shown. No rotation is applied to the data. Details are described and analyzed in Paper~I. The cross-section for day 44 is presented in Fig.~4 in Paper~I. 

Animation~D2. Synthetic soft ($0.5$-$2\un{keV}$) and hard ($2$-$10\un{keV}$) X-ray images with absorption included. Details are described and analyzed in Paper~I. Some snapshots are presented in Fig.~5 in Paper I.

Animation~D3a and D3b. Synthetic \g-ray images for the photon energy range $0.1-300\un{GeV}$ without (D3a) and with (D3b) absorption. The hadronic component (left), leptonic component (middle), and sum of the two (right) are shown. 
Details are described and analyzed in Paper~II. Some snapshots for the hadronic component are presented in Fig.~6 in Paper~II, for an orbital inclination of $35\degr$.

Animations~D4a and D4b. Synthetic radio images at 1 GHz without (Animation~D4a) and with (Animation~D4b) absorption. The thermal free-free component (on the left), the non-thermal synchrotron component (in the middle), and the sum of the two (on the right) are shown. 

Animations~D5a and D5b. The same as D4a and D4b at 45 GHz.

\end{appendix}  

\end{document}